\documentclass[letterpaper,twocolumn,10pt]{article}
\usepackage{usenix-2020-09}

\usepackage{tikz}

\usepackage{amsmath,amsfonts,amsthm}
\usepackage{algorithm,algorithmic,listings}
\usepackage{xspace}
\usepackage{enumitem}
\usepackage{tikz-uml}
\usepackage{microtype}
\usepackage{longtable}
\usepackage[numbers]{natbib}
\usepackage{authblk}
\usepackage{enumitem}
\usepackage{multirow}
\usepackage{graphicx}
\usepackage{amssymb}
\usepackage{pifont}
\usepackage{tcolorbox}% http://ctan.org/pkg/tcolorbox
\definecolor{mycolor}{rgb}{0.122, 0.435, 0.698}% Rule colour
\definecolor{gray1}{gray}{0.3}

\definecolor{darkgreen}{rgb}{0.0, 0.5, 0.0}
\definecolor{darkred}{rgb}{0.82, 0.1, 0.26}

\newcommand{\result}[1]{%
\begin{tcolorbox}[colframe=mycolor,boxrule=0.5pt,arc=4pt,
      left=6pt,right=6pt,top=6pt,bottom=6pt,boxsep=0pt,width=\columnwidth]%
      {#1}
\end{tcolorbox}%
}

\definecolor{red}{rgb}{0.6,0,0} 
\definecolor{blue}{rgb}{0,0,0.6}
\definecolor{green}{rgb}{0,0.8,0}
\definecolor{cyan}{rgb}{0.0,0.6,0.6}

\newcommand{\aflnet}{\textsc{AFLNet}\xspace}

\newcommand{\csfuzz}{\textsc{SGFuzz}\xspace}
\newcommand{\scx}{$\textsc{SGFuzz}_\text{x}$\xspace}
\newcommand{\scy}{$\textsc{SGFuzz}_\text{y}$\xspace}
\newcommand{\lf}{\textsc{LibFuzzer}\xspace}
\newcommand{\hf}{\textsc{Honggfuzz}\xspace}
\newcommand{\ijon}{\textsc{IJON}\xspace}
\newcommand{\todo}[1]{}
\renewcommand{\todo}[1]{{\color{red} TODO: {#1}}}

\makeatletter
\renewcommand\AB@affilsepx{, \protect\Affilfont}
\makeatother

\title{Stateful Greybox Fuzzing}
\author[1]{Jinsheng Ba}
\author[2,3]{\rm Marcel B{\"o}hme}
\author[2]{\rm Zahra Mirzamomen}
\author[1]{\rm Abhik Roychoudhury}
\affil[1]{National University of Singapore}
\affil[2]{Monash University}
\affil[3]{MPI-SP}

\begin{document}
%-------------------------------------------------------------------------------

\date{}

\maketitle

%-------------------------------------------------------------------------------
\begin{abstract}
%-------------------------------------------------------------------------------

Many protocol implementations are reactive systems, where the protocol process is in continuous interaction with other processes and the environment. If a bug can be exposed only in a certain state, a fuzzer needs to provide a specific sequence of events as inputs that would take protocol into this state before the bug is manifested. We call these bugs as ``stateful" bugs. Usually, when we are testing a protocol implementation, we do not have a detailed formal specification of the protocol to rely upon. Without knowledge of the protocol, it is inherently difficult for a fuzzer to discover such stateful bugs. A \emph{key challenge} then is to cover the state space \emph{without} an explicit specification of the protocol.
Finding stateful bugs in protocol implementations would thus involve partially uncovering the state space of the protocol. Fuzzing stateful software systems would need to incorporate strategies for state identification. Such state identification may follow from manual guidance, or from automatic  analysis.

In this work, we posit that manual annotations for state identification can be avoided for stateful protocol fuzzing.  Specifically, we rely on a programmatic intuition that the state variables used in protocol implementations often appear in {\tt enum} type variables whose values (the state names) come from named constants. In our analysis of the Top-50 most widely used open-source protocol implementations, we found that every implementation uses state variables that are assigned named constants (with easy to comprehend names such as INIT, READY) to represent the current state. In this work, we propose to automatically identify such state variables and track the sequence of values assigned to them during fuzzing to produce a "map" of the explored state space.

Our experiments confirm that our stateful fuzzer discovers stateful bugs twice as fast as the baseline greybox fuzzer that we extended. Starting from the initial state, our fuzzer exercises one order of magnitude more state/transition sequences and covers code two times faster than the baseline fuzzer. Several zero-day bugs in prominent protocol implementations were found by our fuzzer, and 8 CVEs have been assigned.

\end{abstract}

\begin{figure}[h]
    \centering
    \includegraphics[width=\columnwidth]{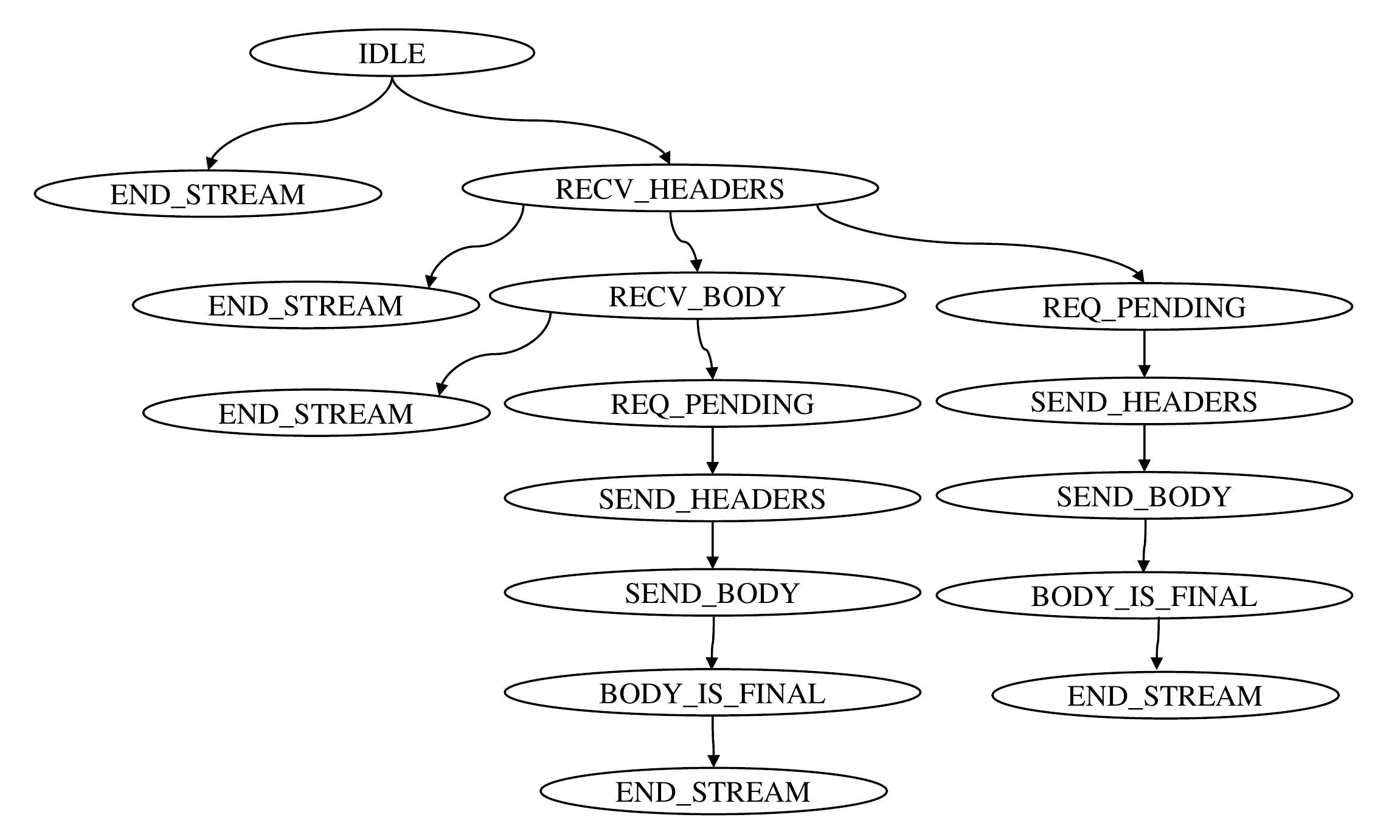}
    \caption{Dynamically constructed state transition tree (STT) for an HTTP2 protocol implementation (H2O).}
    \label{fig:stt}
    %\vspace*{-0.2in}
\end{figure}

\section{Introduction}

Protocol implementations are stateful software systems that require input messages to be sent in a certain expected order. Conventional greybox fuzzing approaches consider structured program inputs, which are subject to random mutations for exploring the space of program inputs. However, for protocols, the input needed to trigger a behavior is a \emph{sequence} of messages, events, or actions. Thus for exposing corner cases or bugs in such systems, which we call stateful bugs, the fuzzing technique needs to explore an unknown state space by constructing and navigating suitable event sequences. We design, implement and evaluate such a fuzzer in this paper.

Existing coverage-guided greybox fuzzers (CGF) \cite{afl,libfuzzer} have been very successful at finding bugs, but they cannot effectively deal with a state space. For instance, the HTTP2 protocol requires a HEADER frame to be sent before the DATA frame. In the H2O implementation (\autoref{fig:stt}), a state variable is \emph{set} to a certain value when the HEADER frame is received. The same state variable is \emph{checked} when the DATA frame is received. Suppose, there is a bug in the response to a correctly processed DATA frame. As there is a specific sequence of input messages needed to reveal the bug, we call this a \emph{stateful bug}. Suppose also that the first generated input sends a DATA frame and then a HEADER frame. Clearly, the DATA frame cannot be processed; yet the code for the HEADER frame handler is already covered. A coverage-guided fuzzer has no information on the traversed state space. It is oblivious to the fact that the correct processing of the DATA frame is "unlocked" only when the requisite state is visited first (during the processing of the HEADER frame). 

The \emph{main objective} of a \emph{stateful greybox fuzzer} should be to chart out and efficiently explore an unknown state space to discover stateful bugs. To this end, a stateful fuzzer (i)~constructs a lightweight abstraction of the observed state space from the state feedback---in our case from the sequence of observed state variable values, and (ii)~navigates this state space abstraction to maximize the probability of visiting unobserved states.
In line with most existing efforts in greybox fuzzing, we assume that \emph{no} external information is available, neither in the form of protocol specifications nor in the form of human annotations \cite{ijonn}. A stateful greybox fuzzer is pointed to the protocol implementation, fuzzes it, and reports any bugs that it finds. So, how can we enable a greybox fuzzer to effectively navigate an unknown state space?

\emph{Identifying state space}. In this paper, we argue that protocols are often explicitly encoded using \emph{state variables} that are assigned and compared to named constants.\footnote{We found that this observation holds for all of the top-50 most widely used protocol implementations as well as for every stateful protocol in the ProFuzzbench protocol testing benchmark \cite{profuzzbench}.}
State variables allow developers to implement state transitions by assigning the corresponding named constant, and to implement state-based program logic as \texttt{if}- or \texttt{switch}-statements.
More specifically, using pattern matching, we identify state variables using enumerated types (\texttt{enum}s). An \emph{enumerated type} is a group of named constants that specifies all possible values for a variable of that type.  Our \emph{instrumentation} injects a call to our runtime at every program location where a state variable is assigned to a new value. Our \emph{runtime} efficiently constructs the state transition tree (STT). The \emph{STT} captures the sequence of values assigned to state variables across all fuzzer-generated input sequences, and as a global data structure, it is shared with the fuzzer. An example of the STT constructed for the H2O implementation of the HTTP2 protocol is shown in \autoref{fig:stt}. Our \emph{in-memory fuzzer} uses the STT to steer the generated input sequences towards under-explored parts of the state space.

\emph{Stateful greybox fuzzing}. We discuss several heuristics to increase the coverage of the state space via greybox fuzzing. First, we propose to add generated inputs to the seed corpus that exercise new nodes in the STT. As we will demonstrate, code coverage alone is insufficient to capture the order across different requests. Instead, we should capture the states in which the code is covered. Hence we argue, adding inputs that discover a new node in the STT facilitates a better coverage of the state space. Secondly, we propose to focus on the seeds which traverse the rarely visited nodes of the STT or whose offsprings are more likely to take a different path through the STT. We hypothesize that these heuristics will help the fuzzer to explore the state space. Finally, we propose an approach to focus particularly on the bytes in the seed, whose mutations trigger new nodes in the STT. The state space corresponding to the newly added nodes can be efficiently explored by mutating these bytes.

\emph{Results}. We implemented our stateful greybox fuzzing approach into \lf \cite{libfuzzer} and call our tool \csfuzz (Stateful Greybox Fuzzer). We evaluated \csfuzz against \lf and \aflnet on eight widely used protocol implementations. 
The state sequence for an input is determined by the sequence of values assigned to the state variables during the execution of the protocol implementation. In our experiments, starting from the initial state, our stateful fuzzer exercises 33x more state sequences than \lf 15x more than \ijon, and 260x more than \aflnet. 
Observing that some code can be covered only in certain states, we found that \csfuzz achieves the same branch coverage more than 2x faster than \lf. 
By reproducing existing stateful bugs, we also found \csfuzz exposes stateful bugs about twice as fast as \lf and \ijon, more than 155x faster than \aflnet. 
Most of all, \csfuzz found 12 new bugs in widely used stateful systems, 8 of which were assigned CVEs. 
Our analysis further shows that stateful bugs are prevalent: every four in five bugs reported among our subjects are stateful.
We have made our data set publicly available for review and will make an open-source release of our fuzzer to researchers and practitioners upon publication of the work.

\vspace{0.2cm}
\noindent
In summary, we make the four key contributions:
\begin{itemize}[itemsep=0cm]
    \item We propose an automatic method to identify and capture the explored state space of a protocol implementation.
    \item We present the design and implementation of \csfuzz, a stateful greybox fuzzer that found 12 new bugs in widely-used and well-fuzzed programs.
    \item We conduct and present a comprehensive evaluation of \csfuzz against the state-of-the-art and the baseline stateful greybox fuzzers.
    \item We make all data, scripts, and tools publicly available to facilitate reproducibility.
\end{itemize}

\section{Motivating Example}\label{sec:motivating}
We believe that current greybox or stateful fuzzers are ineffective in detecting stateful bugs. We use the stream state machine of HTTP2 protocol and the H2O implementation to explain the main reasons for their inefficiency to motivate our approach for stateful greybox fuzzing.

\begin{figure}
    \centering
    \includegraphics[width=0.8\columnwidth]{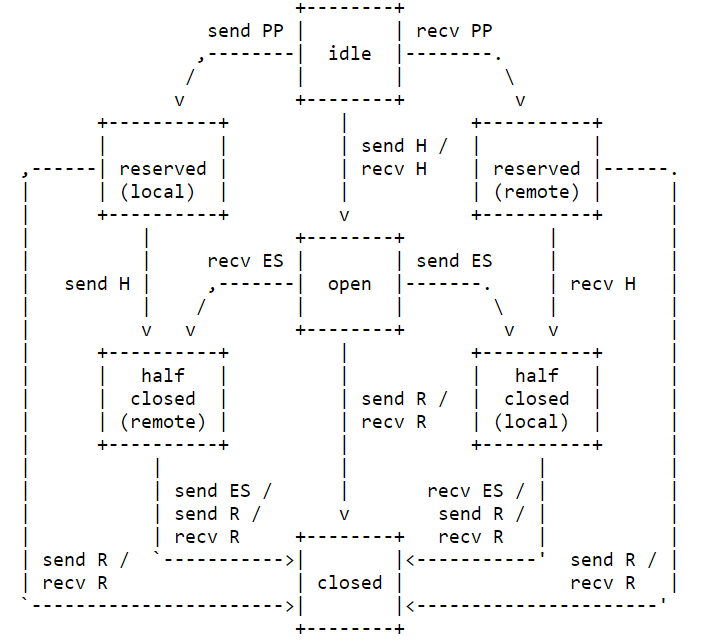}
    \caption{The HTTP2 stream state machine as per RFC 7540.}
    \label{fig:http2SM}
\end{figure}

\subsection{The HTTP2 Protocol}
HTTP2 is a multiplexing protocol, in which each HTTP request is split into several frames corresponding to the HTTP header, HTTP body, and so on. A \emph{frame} is the minimal data unit that is passed along the HTTP2 protocol. For instance, an HTTP2 request message may be split into a header frame and a few data frames. All frames belonging to the same HTTP request constitute a \emph{stream}, having to be processed by the server in order. The HTTP2 protocol defines the \emph{stream state machine}, which regulates the processing order of frames. When different frames are being processed, the server visits different states, e.g., the ``receiving header'' state when it receives the header frame and the ``receiving body'' state when receiving data frames.

\autoref{fig:http2SM} shows the HTTP2 stream state machine as per the official specification (RFC 7540)\footnote{\url{https://datatracker.ietf.org/doc/html/rfc7540}}. In the initial \emph{idle state}, the current or remote participant can send / receive a \emph{push promise} frame (PP) or a header frame (H). A push promise frame simply marks a stream as \emph{reserved} while a header frame marks the stream as \emph{open}. In the open state, all frames are processed including \emph{data} frames. An \emph{end stream} (ES) frame will mark an open stream as \emph{half-closed} and a half-closed stream as \emph{closed}. In any state, if a \emph{reset} (R) frame is sent / received, the stream is marked as closed.

\subsection{The H2O Implementation}\label{sec:h2oimpl}
H2O is a new generation HTTP server that is very fast. H2O supports the HTTP, HTTP2, and HTTP3 protocols and is written in C programming language. We focus on HTTP2 implementation.

\emph{State Variables}. H2O implementation uses a much \emph{finer-grained HTTP2 stream state machine} than specified in \autoref{fig:http2SM}. Specifically, H2O tracks the different stages in which frames are received and responses are sent. H2O uses a dedicated variable to store the current protocol state (i.e., a state variable; here the enumeration variable \texttt{stream->state}). The eight implemented HTTP2 stream states are defined as \emph{enumerated type} with the following named constants:
{\small
\begin{itemize}[itemsep=0.05cm,leftmargin=0.32cm]
    \item \textbf{State~(0).} \texttt{H2O\_HTTP2\_STREAM\_STATE\_IDLE}
    \item \textbf{State~(1).} \texttt{H2O\_HTTP2\_STREAM\_STATE\_RECV\_HEADERS}
    \item \textbf{State~(2).} \texttt{H2O\_HTTP2\_STREAM\_STATE\_RECV\_BODY}
    \item \textbf{State~(3).} \texttt{H2O\_HTTP2\_STREAM\_STATE\_REQ\_PENDING}
    \item \textbf{State~(4).} \texttt{H2O\_HTTP2\_STREAM\_STATE\_SEND\_HEADERS}
    \item \textbf{State~(5).} \texttt{H2O\_HTTP2\_STREAM\_STATE\_SEND\_BODY}
    \item \textbf{State~(6).} \texttt{H2O\_HTTP2\_STREAM\_STATE\_SEND\_BODY\_IS\_FINAL}
    \item \textbf{State~(7).} \texttt{H2O\_HTTP2\_STREAM\_STATE\_END\_STREAM}
\end{itemize}
}
At the start, the server is in the idle State~(0) and waits for incoming requests. Depending on the frames received within the stream, the following states are visited: receiving request header~(1), request body~(2), sending response header~(4), response body~(5), and end of stream~(7). State~(3) is an intermediate state in which an incoming frame has not been assigned a handler, yet. State~(6) is reached only when the client has indicated the end of the stream but there are still pending frames to be sent from the server.

\result{H2O uses {\em named constants} to keep track of the current HTTP2 protocol state. We use this insight to identify states.}

\subsection{Challenges of Fuzzing Stateful Software}
\begin{figure}[h]
    \centering
    \includegraphics[width=0.9\columnwidth]{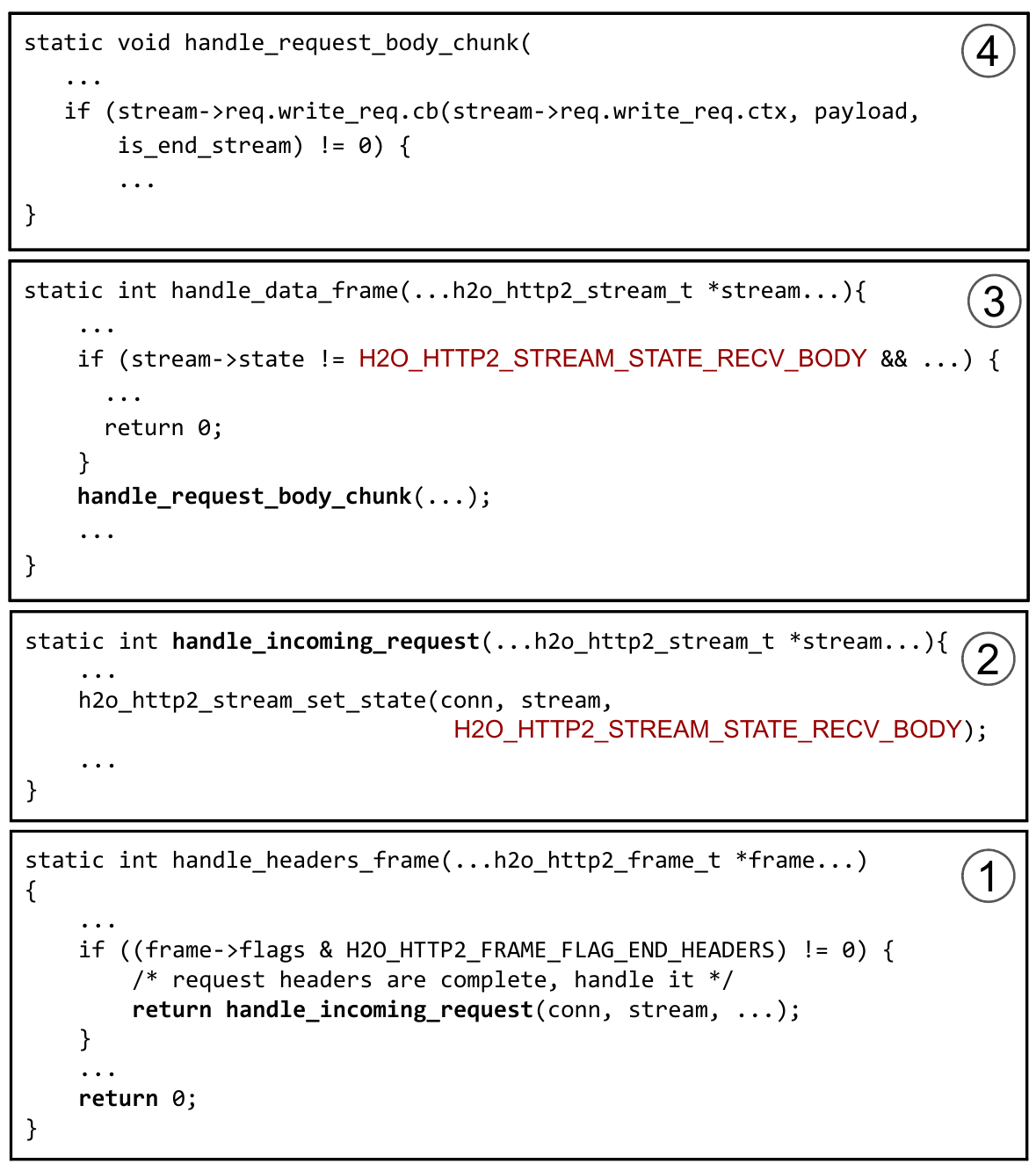}
    \caption{H2O handlers for header and data frames. A handler processes inputs as they arrive. Depending on the sequence in which header and data frames arrive, the value of the state variable is set and checked. Certain code is reachable only if the state variable carries a certain value.}
    \label{fig:code_snippet}
\end{figure}

If the protocol specification is unavailable, how can we automatically figure out the conditions required to exercise the state-dependent code? We make the following observations.

{\em Uninformative Response Codes}. The observable response code is unrelated to the current HTTP2 stream state. This renders ineffective the state-of-the-art fuzzer AFLNet \cite{aflnet} which depends on the response code to identify the current~state.

{\em Coverage is insufficient}. In Figure~\ref{fig:code_snippet}, the function \texttt{handle\-\_request\_body\_chunk}  \textcircled{4} is executed only if the current value of \texttt{stream->state} is \texttt{..RECV\_BODY} when the data frame is handled \textcircled{3}. However, \texttt{stream->state} is set to \texttt{..RECV\-\_BODY} \textcircled{2} only when the handler for the header frame is called beforehand with a valid header frame \textcircled{1}. Code coverage does not capture the required ordering. It is possible to cover both handler functions (\textcircled{1} and \textcircled{3}) by sending a data frame first and a header frame next. However, this sequence of frames cannot be easily modified to cover the function \texttt{handle\-\_request\_body\_chunk}  \textcircled{4}. On the contrary, if the header frame is sent first and the data frame next, the same handler functions would be covered, but this latter sequence can be modified much more easily to cover the function \texttt{handle\-\_request\_body\_chunk}  \textcircled{4} (if it was not already). For existing coverage-guided fuzzers, there would be no distinction between these two scenarios in terms of coverage\footnote{As both handlers are called in a loop, there does not exist a direct edge between both handlers that could be covered.}.
 
\result{By adding generated message sequences to the corpus that exercise a new sequence of states (i.e., add a new node to the state transition tree), we can capture both scenarios.} 
\section{Methodology}\label{sec:method}
Our main objective is to capture the protocol state space without any knowledge of the protocol. Our key observation is that the current protocol states are often explicitly represented using so-called state variables.

\subsection{Offline State Variable Identification}

Our first step is to automatically identify state variables in the program code.
If the reader is tasked to implement a stateful protocol that contains explicitly named protocol states and a well-specified protocol logic that proceeds based on the current protocol state, how would the reader keep track of the current protocol state in his implementation? 
We examined the top-50 most widely used open-source implementations (cf. \autoref{appendix:top50})\footnote{We also confirmed our observation for all stateful protocol implementations in the ProFuzzBench protocol fuzzing benchmark \cite{profuzzbench}.} of stateful protocols as indicated in a search on Shodan\footnote{Shodan is a search engine for various types of servers connected to the internet: \url{https://www.shodan.io/}.} or by Github stars. We found that all of them use named constants assigned to special state variables to keep track of the current state. Of those, 44 use \texttt{enum} type while 6 use \texttt{\#define} statements to define the named constants.
Unlike the state-of-the-art \cite{aflnet}, which requires manual effort to write protocol-specific drivers that extract state information from the server responses, our state variable identification is entirely automatic.

An example of a state variable is given in \autoref{sec:motivating}. Using state variables, developers of protocol implementations can maintain a direct mapping between protocol states given in the protocol specification (cf. \autoref{fig:http2SM}) and the protocol states in the implementation (cf. \autoref{sec:h2oimpl}). State transitions are implemented by assigning another named constant to the state variable. Protocol logic that is based on the current state can be implemented as switch-statements or if-conditions where the current state variable value guards the corresponding state-based protocol logic (c.f. \autoref{fig:code_snippet}).

To identify state variables, we look for all variables of enumerated type (\texttt{enum} in C/C++). An \emph{enumerated type} is a list of named constants used in computer programming to map a set of names to numeric values. Variables of enumerated type can only be assigned constants from the specified list of named constants. In our case, this list represents all states that the state variable can represent (e.g., \texttt{IDLE} to \texttt{END\_STREAM} in \autoref{sec:h2oimpl}). Specifically, we use regular expressions to automatically extract all definitions of enumerated types and then use the definitions to return the list of all enum variables that have been assigned at least once.

As we can see in \autoref{fig:enumtype}, not all variables of enumerated type need to be state variables. The second category is enumerated types that represent all possible response or error codes. The third category is enumerated types that represent all possible configuration options for a configuration variable. However, in practice, our heuristic remains very effective. Response and error code variables are already indirectly used for state identification by the state-of-the-art \aflnet. Configuration variables are usually assigned once and only when the server starts (i.e., before we start recording the state transitions for each message sequence), or at the beginning of the session. If we record configuration variables at the beginning of the sessions, they appear at the beginning of each state transition sequence and will never be identified as “rare” states (which we focus on in \autoref{sec:power}).

\begin{table*}\footnotesize
\begin{tabular}{@{}|p{0.11\textwidth} | p{0.1\textwidth} | p{0.71\textwidth}|@{}}\hline
\textbf{Enum. Type} & \textbf{Example(s)} & \textbf{Explanation}\\\hline
State variables & \texttt{stream->state}, \texttt{sender\_state} & Explicit representation of the protocol states. These are the enum-type variables, the values of which we aim to record in the state transition tree. In H20, 22 of 43 enum-type variables are actual state variables.\\\hline
Error or response codes & \texttt{error}, \texttt{status} & Response codes or error codes are explicitly defined in the protocol and provide quick information to the client about the processing of the last message. These enum-type variables correspond to the existing response-code-based state identification in \aflnet. In H2O, 2 of 43 enum-type variables are response or error codes.\\\hline
Configuration variables & \texttt{priority}, \texttt{run\_mode} & These variables represent the concrete choices from a set of configuration options. Those options are read either from a configuration file or the command line. 
In H2O, 19 of 43 enum-type variables are configuration variables.\\\hline
\end{tabular}
\caption{Kinds of variables with enumerated types and their impact on the state transition tree.}
\label{fig:enumtype}
\end{table*}

\subsection{State Transition Tree Data Structure}\label{sec:inference}
To capture the protocol state transitions exercised by generated sequences of inputs, we monitor the sequences of values assigned to state variables. All observed sequences of values across all state variables in all threads are stored in the same state transition tree, a data structure that represents the entire state space that has been explored by the fuzzer. An example is shown in \autoref{fig:stt}. 

\emph{State Transition Tree}. A node in the state transition tree represents the value of a state variable during program execution. Each node has only one parent node and zero or more children nodes. If there is only one state variable, the parent and the children nodes of a specific node respectively represent the state variable's value before and after the creation of that node. If there are multiple state variables, a node's parent or children nodes can represent values of different state variables. Each edge from a node to its children represents a value change of any state variable, indicating a state transition. From a global perspective, the tree has only one root node which represents the initial state, and each path from the root node to a leaf node represents a unique state transition sequence during program execution on an input.

We use the motivating example from \autoref{sec:motivating} to explain the state transition tree (STT). \autoref{fig:stt} shows the STT for the \texttt{stream->state} state variable after getting five execution traces. The root node is the idle State~(0). After getting valid requests in stipulated order, the following nodes are recorded: receiving request header State~(1), request body State~(2), sending response header State~(4), response body State~(5), and end of stream State~(7). If any request has malformed data, the state directly transitions to the end of stream State~(7) as shown in the left three branches. Another situation is that the request body is empty. In this situation, it will go through the right branch,  bypassing the receiving request body State~(2). 

\begin{figure}
    \centering
    \includegraphics[width=0.8\columnwidth]{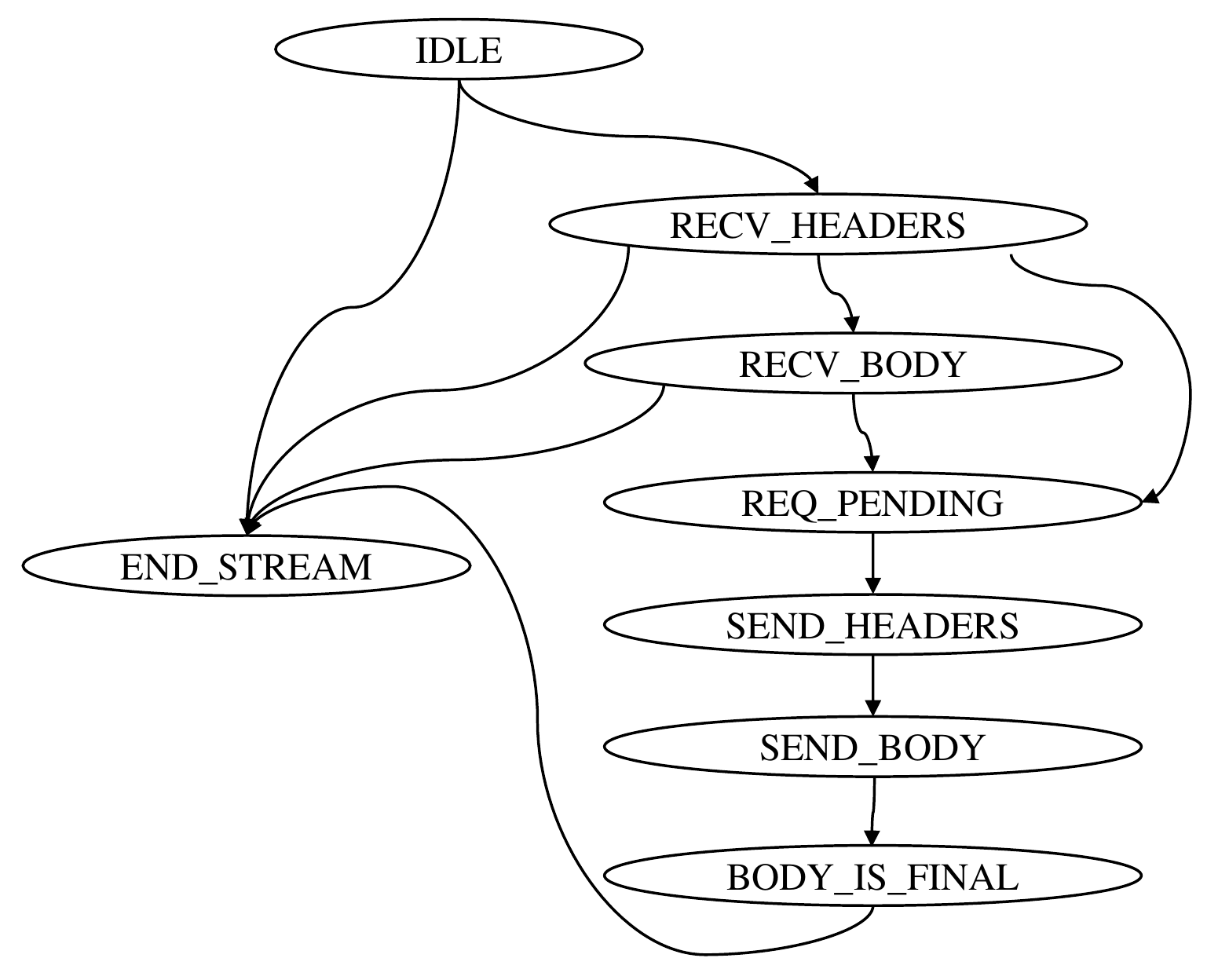}
    \caption{\em The state machine extracted from the State Transition Tree of H2O (node name prefixes omitted).}
    \label{fig:sm}
\end{figure}

\autoref{fig:sm} shows a compact representation of the automatically constructed STT as a directed graph. Visual comparison to the official state machine in \autoref{fig:http2SM} reveals a striking similarity with the specified protocol. Specifically, to construct the directed graph, we merge all nodes with the same values of enumeration variables, such as merging all leaf nodes which have the same State~(7). Like the official state machine, our extracted state machine also starts at the idle state. After receiving a valid header frame, our extracted state machine transitions to the receiving request header State~(1) while the official state machine transitions to the open state. Next, when traversing States~(2)(3)(4)(5), the official state machine is still in the same open state. Finally, when the connection is ready to close, our extracted state machine transitions to State~(6) and State~(7) which corresponds to the half-closed (local) and closed states in the official state machine, respectively. Therefore, we not only accurately recover parts of the official state machine, using a simple but highly effective approach, but we also get a more fine-grained state machine as it is actually implemented. Together with our greybox fuzzing algorithm, we aim to gradually recover more and more states on the fly.

\subsection{State Transition Tree Construction}\label{sec:construct}
To construct the state transition tree (STT), we automatically inject a runtime into the program binary during compilation. The runtime gradually constructs the STT along with program executions on the inputs generated during fuzzing campaign.

\emph{Compile-time instrumentation}. We developed a Python script that injects a call to our runtime before every assignment of a named constant to a state variable in the protocol implementation. Given the list of state variables extracted during our offline state variable identification, we use regular expressions to find all instructions in the code where the state variables are assigned new values. The injected function calls pass the state variable names and the value of the named constant to our runtime. To make the instrumentation more flexible and accurate, we also provide an option to allow users to block some variables extracted from offline state variable identification. We did not choose LLVM to instrument the protocol implementation because \texttt{enum}-values will be replaced by integer-type constant values in the LLVM intermediate language (LLVM-IR).

\emph{Runtime derivation of State Transition Tree}. Across all executions of a protocol implementation during a fuzzing campaign, the runtime uses the information passed by the injected calls, to gradually construct the STT. The STT data structure is implemented in the runtime component with a pointer that points to the last visited node (initially root).
Each node in the STT is distinguished by the variable name and its corresponding value, passed from the instrumented calls to the runtime. Once called, the runtime checks whether a matching node already exists among the child nodes of the last visited node. If not found, a new child node is created for it. Then the last-visited-node pointer is set to point to the already existing or the newly created child node, whichever applies. After execution of the program on each input, the last-visited-node pointer is reset to the root node waiting for the next execution. To support multi-threading, which is common for protocol implementations, we leverage a mutex to protect the code for updating the state transition tree from concurrency issues. To avoid that the STT grows to an unreasonable size, we provide an option to limit the maximum number of state repetitions along with a single execution. Whenever a state is repeated more often than allowed (along a path in the STT), subsequent states are ignored (and that particular path is truncated). This allows us to efficiently manage the state transitions as a tree, rather than a graph, during the fuzzing campaign. In our experiments,we chose the same threshold for all subjects, generally so that the number of nodes does not exceed 10k nodes.

\subsection{Handling Implicit States}\label{sec:implicit}
In addition to the explicit states that are captured using state variables, like for any software system, there exist implicit states: databases and file systems are changed; some memory is allocated but never freed. While explicit states are reset whenever the connection is closed after a message exchange, implicit states aggregate over time and survive various connection resets.\footnote{A connection includes multiple requests and responses.} The STT can only keep track of explicit states but not of such implicit states. To explore the implicit state space, we let the implicit states aggregate without reset. This allows us to find other types of stateful bugs.

Our in-memory stateful greybox fuzzer lives in the same process as the protocol implementation. Hence, the protocol implementation under test does not need to be restarted for every message exchange. This gives a substantial performance boost but also allows us to aggregate implicit states across message exchanges.
To facilitate reproducing this type of bug, we enable an option to save the list of all the inputs on which the program has been executed. After getting a crash report from the fuzzer, we gradually minimize the list until a minimal list triggers this crash. This minimal list can be used for further manual analysis.
\section{Stateful Greybox Fuzzing Algorithm}\label{sec:fuzzing}
To efficiently explore the state space, we propose to use the State Transition Tree (STT), described in section (\ref{sec:inference}), as a guide for the stateful greybox fuzzer.  \csfuzz incrementally constructs the STT at runtime and employs it as follows:
\begin{enumerate}[itemsep=0cm]
  \item Generated inputs that increase coverage of the STT are added to the seed corpus.
  \item Within the seed corpus, \csfuzz prioritizes the seeds that %execute rarely executed STT nodes which 
  have a greater potential to increase the coverage of the STT.
  \item Within the chosen seed, \csfuzz prioritizes bytes associated with newly added STT nodes which have more potential to increase the coverage of the STT.
\end{enumerate}

\renewcommand{\algorithmicrequire}{\textbf{Input:}}
\renewcommand{\algorithmicensure}{\textbf{Output:}}
\begin{algorithm}[h]
\caption{Stateful Greybox Fuzzing}
\label{alg:fuzzing}
\begin{algorithmic}
\REQUIRE {Seed Corpus T}
\STATE {Crashing inputs $T_{x}=\emptyset$}
\FOR {\textbf{each} $t\in T$}
\STATE $E_t=1$ \quad\quad\textcolor{gray}{\emph{// initial energy}}
\ENDFOR
\STATE {\textcolor{red}{{State transition tree $STT=\emptyset$}}}
\REPEAT
\STATE {$t=$\emph{ choose\_next}$(T, E)$}
\STATE $t'=$\textcolor{red}{{ $mutate(t, STT)$}}
\STATE $\textcolor{red}{STT = execute(t', STT)}$
\IF {$t'$ crashes}
\STATE {\textbf{return} $t'$}
\ELSIF {\textcolor{red}{\emph{is\_interesting}($t', STT)$}}
\STATE {add $t'$ to $T$}
\STATE $t' = \textcolor{red}{{identify\_bytes(t', STT)}}$
\FOR {\textbf{each} $t_i\in T$}
  \STATE {$E_{t_{i}}=$\textcolor{red}{\emph{ assign\_energy}$(t_i, STT)$}}
\ENDFOR
\ENDIF
\UNTIL {timeout reached or abort-signal}
\ENSURE {Crashing Input $t_x$}
\end{algorithmic}
\end{algorithm}

Algorithm~\ref{alg:fuzzing} shows the general workflow of our fuzzing algorithm.  We implemented this algorithm into \lf and call our fuzzer \csfuzz (Stateful Greybox Fuzzer). Starting with an initial seed corpus $T$ and until a timeout is reached or the user aborts the campaign, the following steps are repeated. Firstly, \csfuzz samples $t$ according to the energy distribution $E$ which represents the likelihood of each seed in $T$ to be chosen (\emph{choose\_next}). The chosen seed $t$ is then mutated (\emph{mutate}) to generate a new input $t'$ (\autoref{sec:mutation}). Then we execute the protocol implementation on $t'$ and update the STT (\autoref{sec:construct}). If $t'$ causes a crash \csfuzz returns the crashing input, else if $t'$ exercises new code branch or new STT nodes (\emph{is\_interesting}), $t'$ is added to the corpus $T$, and otherwise $t'$ is discarded (\autoref{sec:feedback}). Whenever new STT nodes added, \csfuzz identifies the bytes in input $t'$ (\emph{identify\_bytes}) that contributed to the coverage of the new STT nodes (\autoref{sec:mutation}) and updates the energy $E_{t_i}$ of all seeds $t_i\in T$ in the current corpus (\emph{assign\_energy}) (\autoref{sec:power}). The portions of \lf we changed are highlighted in red in Algorithm~\ref{alg:fuzzing}.

\subsection{State Coverage Feedback}\label{sec:feedback}
Our first heuristic is to add generated inputs to the seed corpus that exercise new nodes in the STT (\emph{is\_interesting} in Alg.~\ref{alg:fuzzing}) as the augmentation of the original branch coverage feedback.
The rationale is to maximize the coverage of the STT data structure to maximize the coverage of the state space. Each new node in the STT represents a new value in the current sequence of state variable values indicating a new state under the current sequence of state transitions. Like in code-coverage-guided greybox fuzzing, inputs that increase coverage are added to the seed corpus, as well.

\subsection{Energy Schedule}\label{sec:power}

Our second heuristic is to assign more energy to seeds that have a greater potential to lead to new states (\emph{assign\_energy} in Alg.~\ref{alg:fuzzing}). Compared to existing works\cite{aflfast, tai2020ecofuzz} that prioritize inputs that execute rarely executed code blocks, our algorithm additionally assigns more energy to the seeds that execute the valid state transitions which correspond to the valid protocol behaviors. 

Firstly, we assign more energy to the seeds that traverse the rarely visited nodes in the STT. We set up an attribute called \emph{hit count} for each node in the STT to record the number of inputs that traverse that node. We expect that STT nodes with low \emph{hit count} have more unexplored states in their "neighborhood". Given the STT, we consider a node $n\in STT$ to be \emph{rare} if it has a below-average \emph{hit count}. 
\begin{equation}\label{equ:rare}
\begin{split}
    rare(n) = \begin{cases}
    1 & \text{ if } hits(n) < \frac{\sum_{n'\in STT}hits(n')}{|STT|} \\
    0 & \text{ otherwise}
  \end{cases}
\end{split}
\end{equation}
where $hits(n)$ returns the \emph{hit count} of node $n$.

Let $path(t)$ represent the set of nodes that are traversed by input $t$ in STT.
Given an input $t$ in corpus $T$, our power schedule assigns energy $e_1(t)$ of $t$ as the proportion of rare nodes that are visited during program execution on $t$ added to the original energy of $t$:
\begin{equation}\label{equ:power_rare}
    e_1(t) = e'(t) +  e'(t)\frac{\sum_{n\in path(t)}rare(n)}{|path(t)|}
\end{equation}
where $e'(t)$ is the existing power schedule.

Secondly, we assign more energy to the seeds whose offspring are more likely to take a different path through the STT. By \autoref{equ:power_rare}, all nodes in STT have similar opportunities to be executed. However, not all state transitions are equally important, and valid state transitions are more attractive. Our intuition is that the valid state transitions, which correspond to the expected protocol behaviors, are usually easy to be mutated to other invalid state transitions, which represent the error handling of the protocol. The change on any byte may incur error handling logic. This is also a challenge of previous stateful greybox fuzzing techniques. Thus, we identify this type of seed and assign more energy to it. 
Given an input $t$ and a set of inputs $t_m$ that are mutated from $t$, we calculate the inverse proportion of its offspring that traverse the same paths as $path(t)$. We extend our power schedule $e_2(t)$ as follows:
\begin{equation}
    e_2(t) = e_1(t) \cdot \frac{|t_m|}{\sum_{c \in t_m}(path(c) == path(t))} 
\end{equation}

Finally, to avoid assigning too much energy, we limit our power schedule $e(t)$ to no more than ten times (empirically works best than other numbers) the original energy $e'(t)$. Given the seed input $t\in T$ in seed corpus $T$, our power schedule assigns the new energy $e(t)$ to $t$. 
\begin{equation}
    e(t) = \min(e_2(t), 10 \cdot e'(t))
\end{equation}

\subsection{Fuzzing Individual Bytes}\label{sec:mutation}
We define a simple heuristic to fuzz parts of the seed first that might have more impact on increasing state (STT) coverage.
When an input seed is chosen to be mutated, its individual bytes to be mutated are selected based on the bytes' priorities. Bytes have higher priority if they are deemed to have more impact on the STT's coverage (\emph{mutate} and \emph{identify\_bytes} in Alg.~\ref{alg:fuzzing}). Recent works on input probing \cite{gan2020greyone, you2019profuzzer, aschermann2019redqueen} establish the information flow from particular input bytes to a given sink in a very lightweight manner. We leverage these ideas to explore state space. If program execution on a mutated input triggers new nodes in the STT and hence, that input is causing new unexplored nodes to be visited for the first time, it may have a great potential to increase the coverage of the state space. The idea is to identify the individual bytes in such an input whose mutation has contributed to the increase in the STT's coverage. Once identified, those bytes will be assigned higher priority to be chosen for mutation. It makes fuzzing more efficient by enabling earlier exploration of the "neighbor" states around the newly created nodes that have not been visited, yet.

Given an input $t$ and a mutated input $t'$ derived from $t$, if the program execution on input $t'$ creates new nodes in STT, $t'$ is saved by \csfuzz. Then \csfuzz computes the set difference of bytes $diff(t,t')$ between $t$ and $t'$ and records the set of newly added STT nodes $N$ exercised by $t'$.
We annotate the mutation range as $R_{t'}=diff(t,t')$ if $|N|>0$, or set $R_{t'}$ to the entire seed $t'$ ($R_{t'}=\langle t'\rangle$), otherwise (\emph{identify\_bytes} in Alg.~\autoref{alg:fuzzing}).
Later, when $t'$ is chosen to be mutated for fuzzing ($mutate$ in Alg.~\autoref{alg:fuzzing}), we only mutate the bytes under the range $R_{t'}$. If it does not bring any interesting behavior ($is\_interesting$ in Alg.~\autoref{alg:fuzzing}), we gradually enlarge $R_{t'}$ until it spans the entire seed $t'$. This helps us build an efficient fuzz campaign.

\section{Experimental Setup}

\subsection{Research Questions}
Our main hypothesis is that stateful greybox fuzzing allows for more effective fuzzing of stateful programs, such as protocol implementations. We investigated this hypothesis along with the following research questions.
\begin{description}[itemsep=0.1cm]
  \item[RQ.1 (State Transition Coverage)] How many more different sequences of state transitions does \csfuzz exercise? By mapping out the state space, \csfuzz aims to penetrate deeper into the program's state space. The number of different sequences of states exercised by a fuzzer indicates how deep the fuzzer reaches into the state space.  
  \item[RQ.2 (Branch Coverage)] How much more branch coverage does \csfuzz achieve? We cannot find bugs in code that is not covered.
  \item[RQ.3 (State Identification Effectiveness)] Identifying state variables, what is the false positive / false negative rate? How does it affect the mapping of the state space (STT)?
  \item[RQ.4 (Bug Finding Effectiveness)]  How much faster does \csfuzz expose stateful bugs in our benchmark?
  \item[RQ.5 (New Bugs)] How many previously unknown bugs can \csfuzz find in widely used stateful programs? How many are stateful?
  
\end{description}

\subsection{Subject Programs Studied}\label{sec:benchmarks}

\begin{table}
    \centering
    \footnotesize
    \begin{tabular}{@{ \ }r@{ \ }|@{ \ }r@{ \ }r@{ \ }r@{ \ }r@{ \ }r@{ \ }}
        \textbf{Subject} & \textbf{Protocol} &\textbf{Fuzz Driver} & \textbf{Commit} & \textbf{Size} & \textbf{Share}\\\hline
        H2O & HTTP & h2o-fuzzer-http2 & \texttt{1e7344} & 337k LoC& 12k IPs \\
        MbedTLS & SSL/TLS &dtlsserver & \texttt{e483a7} & 138k LoC&  8m IPs\\
        Curl & Several & curl\_fuzzer & \texttt{aab3a7} & 202k LoC& - \\
        Gstreamer & Custom & gst-discoverer & \texttt{44bdad} & 235k LoC& 15k IPs\\
        OpenSSL & SSL/TLS & netdriver & \texttt{1e08f3} & 673k LoC& >10m IPs\\
        Live555 & RTSP & netdriver & 21.Aug'08 & 17k LoC& 12k IPs\\
        OwnTone & DAAP & netdriver & \texttt{774d7c} & 37k LoC& 10k IPs\\
        DCMTK & DICOM & netdriver & \texttt{24ebf4} & 38k LoC& 3k IPs
    \end{tabular}
    \caption{Protocol implementations used for comparison.} 
    \label{tab:benchmark}
\end{table}

\emph{Benchmarks}. We constructed our set of subjects from two mature benchmarks.
\textsc{ProFuzzBench}~\cite{profuzzbench} is a benchmark for stateful fuzzing of network protocols and facilitates comparison to \aflnet \cite{aflnet} (i.e., the current state-of-the-art stateful fuzzer). \textsc{ProFuzzBench} includes a suite of representative open-source network servers for popular protocols.
\textsc{OSS-Fuzz}\footnote{\url{https://google.github.io/oss-fuzz/}} is an open-source continuous fuzzing framework that builds more than 500 open source projects for fuzzing.
\textsc{FuzzBench} \cite{metzman2021fuzzbench} is an open-source fuzzer evaluation infrastructure that automatically conducts fuzzing experiments and produces coverage graphs. \textsc{FuzzBench} supports integrating benchmarks from the \textsc{OSS-Fuzz}.

\emph{Selection Criteria}. We selected 8 real-world target programs as subjects for our experiments as shown in \autoref{tab:benchmark}. We opted for the programs that 1) work on distinct and well-known protocols aiming to provide diverse state machines, 2) are widely used (as indicated by a search on the Shodan), 3) are used in security-critical industries, such as medical and finance, in which any bug can have serious consequences. At the time of writing, they represent the most recent version (cf. \emph{commit} column) of eight programs implementing more than 30 different well-known protocols (curl includes multiple protocol implementations). These programs are widely used across many machines on the internet (cf. \#IPs in the \texttt{Share} column). \textsc{ProFuzzBench} needs socket communication and we use \emph{netdriver}\footnote{\url{https://github.com/google/honggfuzz/tree/master/libhfnetdriver}} to enable it for in-memory fuzzers. Unfortunately, \emph{netdriver} does not support the UDP protocol or the fork system call because a fork will spawn a new process that prevents fuzzers from measuring branch coverage. We selected all programs in \textsc{ProFuzzBench} which are compatible with in-memory fuzzers.

\emph{Selected Subjects}. \texttt{H2O} is a popular web server; \texttt{MbedTLS} and \texttt{OpenSSL} are security-critical encryption libraries; \texttt{Curl}\footnote{\texttt{Curl} implements 26 client-side protocols which Shodan cannot identify.} is a utility to communicate via multiple protocols and implements many state machines; \texttt{Gstreamer} implements a real-time streaming protocol;\footnote{However, the OSS-Fuzz fuzz driver targets the custom protocol which retrieves information about a remote media file.} \texttt{Live555} and \texttt{OwnTone} are libraries for streaming media; and \texttt{DCMTK} is used for the medical data exchange between hospitals (where bugs can be literally fatal). Three quarter of those protocol implementations are running on more than 10 thousand different machines (IPs), according to a recent query of the Shodan search engine.

\subsection{Fuzzers}\label{sec:fuzzers}
We implemented our stateful greybox fuzzing approach into \lf (the most recent version at the time of our experiment) and call it Stateful Greybox Fuzzer (\csfuzz). In our experiments, \csfuzz is compared against \lf as the baseline and \aflnet and \ijon as the state-of-the-art.

\emph{Stateful greybox fuzzer} (\csfuzz).
To implement our algorithm into \lf, we extended the \emph{Fuzzer::RunOne} function with the heuristic described in \autoref{sec:feedback}, changed the \emph{InputCorpus::UpdateCorpusDistribution} function to implement the heuristics discussed in \autoref{sec:power}, and added a filter to the \emph{MutationDispatcher::Mutate} function for the heuristics discussed in \autoref{sec:mutation}. As for the construction of the STT discussed in \autoref{sec:construct}, we implemented a standalone python script for the instrumentation component which is static and does not rely on any compiler infrastructure. The runtime component is embedded in the fuzzer.

\emph{Baseline}.
\lf is in-memory, coverage-guided, evolutionary fuzzer and serves as the baseline fuzzer we used to implement \csfuzz. 

\emph{State-of-the-art}.
\aflnet is the current state-of-the-art stateful fuzzer for protocol implementations that uses the concept of state-feedback. 
\ijon is another state-of-the-art stateful fuzzer which needs manual annotation for states. We used both to evaluate if \csfuzz gains an advantage over the state-of-the-art stateful fuzzer. Because \ijon does not support \textsc{FuzzBench} or \textsc{ProFuzzBench} and to facilitate a fair comparison to our baseline and our implementation, we re-implemented \ijon in \lf and ported its annotation function \texttt{ijon\_push\_state}\footnote{\url{https://github.com/RUB-SysSec/ijon/blob/56ebfe34709dd93f5da7871624ce6eadacc3ae4c/llvm\_mode/afl-llvm-rt.o.c\#L75}} for \ijon-based state-feedback. To maximize the benefit for \ijon, we assume a human annotator would annotate the same state variables as identified by \csfuzz.  \aflnet does not support \emph{FuzzBench}.

\emph{Benchmark Integration}.
All fuzzers were integrated into \textsc{FuzzBench} or \textsc{ProFuzzbench} using equivalent setups and build processes. All fuzzers are provided the same setup, environment, seeds, and resources. For all fuzzers, AddressSanitizer was enabled to detect memory corruption bugs. While \textsc{ProFuzzbench} uses the \texttt{gcov} tool, \textsc{FuzzBench} uses the \texttt{llvm-cov} tool to measure the number of branches covered in a fuzzing campaign in 15-minute intervals. 

\subsection{Experimental Infrastructure}

\emph{Computational Resources}. 
After the integration of \csfuzz and our benchmarks into FuzzBench, we received assistance from the FuzzBench team to run the branch and state transition coverage experiments on the Google Cloud Service \cite{metzman2021fuzzbench}. 
For the other four subjects in the ProFuzzBench, we ran the branch and state transition coverage experiments on our local machine with an Intel Xeon Gold 6258R CPU that has 128 logical cores running at 2.70GHz. The machine runs Ubuntu 20.04.2 LTS and has access to 128GB of main memory.

\emph{Experiment configurations}. 
As for the \emph{initial seed corpus}, we selected one valid input for each subject as the initial seed for RQ1, RQ2, and RQ3, and selected all seeds provided FuzzBench or ProFuzzBench as the initial seed for RQ4 and RQ5. In order to \emph{mitigate the impact of randomness}, we repeated every 23 hours fuzzing campaign 20 times. The 23 hours timeout is the default for the FuzzBench team. To make the results consistent, we use the same time budget. 
\section{Evaluation Results}\label{sec:eva}

\subsection*{RQ.1 State Transition Coverage}

\begin{table}[]
    \setlength{\tabcolsep}{2.7pt}
    \centering
    \scriptsize
    \begin{tabular}{r|cccc|rrr}
        \textbf{Subject} & \textbf{\aflnet} & \textbf{\lf} & \textbf{\ijon} & \textbf{\csfuzz} & \textbf{$\hat A_{12}$} & \textbf{$U$} & \textbf{Factor} \\
        \hline

\multirow{2}{*}{H2O}		& -               & 70.80     & 91.85     & 1849.30     & \multirow{2}{*}{1.00}   & \multirow{2}{*}{<0.01}   & \multirow{2}{*}{26.1} \\
                            & -               & (3.61\%)  & (4.68\%)  & (94.26\%)  \\\hline
\multirow{2}{*}{MbedTLS}	& -               & 22.80     & 32.45     & 50.80       & \multirow{2}{*}{1.00}   & \multirow{2}{*}{<0.01}   & \multirow{2}{*}{2.2} \\
                            & -               & (39.31\%) & (55.95\%) & (87.59\%)  \\\hline
\multirow{2}{*}{Curl}		& -               & 150.25    & 375.75    & 14630.80    & \multirow{2}{*}{1.00}   & \multirow{2}{*}{<0.01}   & \multirow{2}{*}{97.3} \\
                            & -               & (0.98\%)  & (2.45\%)  & (95.55\%)  \\\hline
\multirow{2}{*}{Gstreamer}	& -               & 49.40     & 134.20    & 4067.30     & \multirow{2}{*}{1.00}   & \multirow{2}{*}{<0.01}   & \multirow{2}{*}{82.3} \\
                            & -               & (1.08\%)  & (2.94\%)  & (88.96\%)  \\\hline
\multirow{2}{*}{OpenSSL}	& 13.25           & 23.95     & 29.60     & 33.10       & \multirow{2}{*}{0.81}   & \multirow{2}{*}{<0.01}   & \multirow{2}{*}{1.4} \\
                            & (33.97\%)       & (61.41\%) & (75.90\%) & (84.87\%)  \\\hline
\multirow{2}{*}{Live555}	& 138.27          & 184.15    & 405.3	  & 1162.30     & \multirow{2}{*}{1.00}   & \multirow{2}{*}{<0.01}   & \multirow{2}{*}{6.3} \\
                            & (11.42\%)       & (15.21\%) & (33.75\%) & (95.98\%)  \\\hline
\multirow{2}{*}{OwnTone}	& 1.00            & 46.40     & 426.00	  & 930.15      & \multirow{2}{*}{0.71}   & \multirow{2}{*}{0.03}    & \multirow{2}{*}{20.0} \\
                            & (0.10\%)        & (4.82\%)  & (44.28\%) & (96.69\%)  \\\hline
\multirow{2}{*}{DCMTK}	    & 68.10           & 189.25    & 267.50	  & 6737.05     & \multirow{2}{*}{1.00}   & \multirow{2}{*}{<0.01}   & \multirow{2}{*}{35.6} \\
                            & (0.92\%)        & (2.55\%)  & (3.93\%)  & (90.87\%)  \\\hline
    \multicolumn{5}{c}{} & \multicolumn{3}{r}{\textbf{Avg:} 33.9x}  \\
    \end{tabular}
    \caption{Average state transition coverage for 20 campaigns of 23 hours. The percentage shows this number as a proportion of distinct transition sequences exercised by any run of any fuzzer. Vargha-Delaney ($\hat A_{12}$) measures the magnitude of the difference in transition coverage at 23 hours between \csfuzz and \lf (effect size) while Wilcoxon rank-sum test ($U$) measures the degree to which these results may be explained by randomness (statistical significance). \emph{Factor} shows the factor improvement compared to \lf.}
    \label{tab:transitions}
\end{table}

For all fuzzers, subjects, and campaigns, we measured state transition coverage as the number of paths in the STT which is constructed across the execution of the subject on \emph{all} seeds generated throughout the campaign and from the initial corpus. Each path represents a unique sequence of state transitions during the execution of the subject on an input. If a fuzzer exercises more such succinct sequences, it penetrates deeper into the protocol's state space. 

\emph{Presentation}. \autoref{tab:transitions} shows the average number of state transition sequences for 20 campaigns of 23 hours experiments. To assess the "completeness" of the state space coverage, we counted the number of distinct sequences for any run of any fuzzer and consider this as 100\% for all fuzzers and the average percentage of state transitions exercised by each fuzzer is shown in the brackets. \emph{Factor} represents the factor improvement of \csfuzz to \lf in terms of average \#state sequences. 
Vargha Delaney $\hat A_{12}$ measures the \emph{effect size} and gives the probability that a random fuzzing campaign of \csfuzz achieves more state transition coverage than a random fuzzing campaign of \lf (i.e., $\hat A_{12}>0.5$ means \csfuzz is better). The Wilcoxon rank sum test $U$ is a non-parametric \emph{statistical hypothesis test} to assess whether the coverage differs across both fuzzers. We reject the null hypothesis if $U<0.05$, i.e., \csfuzz outperforms \lf with statistical significance. \aflnet does not support \emph{FuzzBench}. We annotated the same states as in \csfuzz to \ijon which needs manual state annotations. 

\emph{Results}. For all subjects, \csfuzz covers substantially more state transitions than \aflnet, \lf, and \ijon. Compared to baseline \lf, \csfuzz exercises 34x more state transitions on average (20x median) across all eight subjects, owing to the STT guided fuzzing approach. Especially for \emph{Curl}, \emph{Gstreamer}, and \emph{DCMTK}, \csfuzz increases the state transition coverage significantly. Those have much more state variables than others. \csfuzz even executes 97x more state transitions than \lf on \emph{Curl} and 930x more state transitions than \aflnet on \emph{OwnTone}.  An extended discussion about about the contribution of the individual heuristics in \csfuzz can be found in \autoref{sec:sensitivity}.

\aflnet vastly underperforms. We believe the reasons are 1) \aflnet identifies state by a different method (using the response code) and 2) \aflnet restarts the tested program after each generated message sequence which incurs a big performance loss. In contrast, other fuzzers leverage the same in-memory architecture for high throughput. 

\ijon exercises 2.75x more state transitions than \lf on average as it adds state feedback, but \csfuzz still exercises 15.29x more state transitions than \ijon. As a proportion of all sequences exercised in any run of any fuzzer, \csfuzz covers 92\% of state transitions, on average (99\% on the sum of 20 campaigns). This shows that \csfuzz can cover almost all state transitions covered by other fuzzers and more. 1\% sequences are executed by \lf only because of the randomness of fuzzing. 

\lf, as the only stateless fuzzer, performs well only on OpenSSL where the total number of distinct state transition sequence is relatively low (39). We found that \lf explores states mostly near the initial states or end states  (cf. \autoref{appendix:bugprev}). 

For subjects \emph{MbedTLS} and \emph{OpenSSL}, we found that \csfuzz has smaller state transition increase than for the other subjects. For those subjects, the number of distinct sequences explored by any fuzzer is relatively low. For instance, in \emph{OpenSSL}, \emph{st->hand\_state} variable is an enum-type variable used for tracking handshake states of TLS protocol. From state \emph{TLS\-\_ST\-\_SR\-\_CLNT\-\_HELLO}\footnote{\url{https://github.com/openssl/openssl/blob/c74188e/ssl/statem/statem_srvr.c\#L585}}, it needs at least four more conditions to reach state \emph{DTLS\-\_ST\-\_SW\-\_HELLO\-\_VERIFY\-\_REQUEST}, and at least two more conditions to reach state \emph{TLS\-\_ST\-\_OK}. Some of the conditions need a specific external state. For instance, \emph{SSL\-\_IS\-\_DTLS(s)} is true when the traffic comes via UDP protocol rather than TCP. 

\result{On average, \csfuzz covers 33x more sequences of state transitions than \lf, 15x more than \ijon, and 260x more than \aflnet. By tracking the state space via the STT, \csfuzz is able to cover more of the state space.}

\subsection*{RQ.2 Branch Coverage}

\begin{table*}[]
    \setlength{\tabcolsep}{2.2pt}
    \centering\footnotesize
    \begin{tabular}{r|rrrrrrr|rrrrrrr}
\multirow{2}{*}{\textbf{Subject}} & \multicolumn{7}{c|}{\textbf{Branch Coverage}} & \multicolumn{7}{c}{\textbf{ Time-to-Coverage}}\\		
&  \aflnet & \lf & \ijon & \csfuzz & $\hat A_{12}$ & $U$ & Improvement & \aflnet & \lf & \ijon & \csfuzz & $\hat A_{12}$ & $U$ & Factor \\\hline
H2O         & -         & 2879.20   & 2820.25	& 3050.68   & 0.84 & <0.01  & 5.96\%    & -         & 23.00h & 14.04h    & 6.69h    & 0.98  & <0.01 & 3.4           \\
MbedTLS     & -         & 11929.60  & 11885.65	& 12057.68  & 0.58 & 0.43   & 1.07\%    & -         & 23.00h & >23.00h   & 17.86h   & 0.75  & <0.01 & 1.3           \\
Curl        & -         & 19262.16  & 16892.55	& 19942.39  & 0.71 & 0.03   & 3.53\%    & -         & 23.00h & 21.89h    & 12.23h   & 0.95  & <0.01 & 1.9           \\
Gstreamer   & -         & 63280.06  & 60993.58	& 61698.56  & 0.11 & <0.01  & -2.50\%   & -         & 23.00h & >23.00h   & >23.00h  & -     & -     & -             \\
OpenSSL     & 11342.00  & 12463.14  & 12456.00  & 12610.50  & 0.72 & 0.02   & 1.18\%    & >23.00h   & 23.00h & >23.00h   & 10.08h   & 1     & <0.01 & 2.3           \\
Live555     & 2179.87   & 2244.63   & 2214.38   & 2278.10   & 0.73 & 0.02   & 1.49\%    & >23.00h   & 23.00h & >23.00h   & 15.86h   & 0.95  & <0.01 & 1.5           \\
OwnTone     & 991.00    & 2184.64   & 2299.20   & 2312.60   & 0.65 & 0.15   & 5.86\%    & >23.00h   & 23.00h & 6.14h     & 5.04h    & 1     & <0.01 & 4.6           \\
DCMTK       & 5763.00   & 6042.15   & 5997.05   & 6100.50   & 0.77 & <0.01  & 0.97\%    & >23.00h   & 23.00h & >23.00h   & 14.40h   & 0.93  & <0.01 & 1.6           \\

    \hline
    \multicolumn{6}{c}{} & & \textbf{Avg:} 2.20\% & \multicolumn{5}{c}{} & \multicolumn{2}{r}{\textbf{Avg:} 2.3x}  \\
    \end{tabular} 
    \caption{Average branch coverage (left) and average time to coverage (right) for 20 campaigns of 23 hours. The time to coverage measures the time it takes \aflnet, \ijon, and \csfuzz to achieve the same coverage that \lf achieves in 23 hours (0.25h accuracy). Vargha-Delaney ($\hat A_{12}$) measures the magnitude of the difference (effect size) while Wilcoxon rank-sum test ($U$) measures the degree to which these results may be explained by randomness (statistical significance). \emph{Factor} shows the factor improvement compared to \lf.
    }
    \label{tab:coverage}
    \vspace*{-0.1in}
\end{table*}

\emph{Presentation}. \autoref{tab:coverage} shows the branch coverage result after 23 hours. \emph{Improvement} represents the percentage improvement of \csfuzz to \lf based on the average branch coverage. \emph{time-to-coverage} measures the time it takes \aflnet, \ijon, and \csfuzz, respectively to achieve the same coverage that \lf achieves in 23 hours. \emph{Factor} represents the factor improvement of \csfuzz to \lf based on time-to-coverage. $\hat A_{12}$ represents the Vargha Delaney effect size while $U$ represents Wilcoxon signed rank statistical significance. 

\emph{Results}. Although \aflnet also aims to explore the state space of the program, there are substantial inefficiencies that prevent \aflnet from achieving good coverage on most subjects. \ijon performs slightly worse than \lf because it introduces performance cost for state tracing. However, with our algorithm, in general, \csfuzz overcomes the performance cost caused by the STT and outperforms both \lf and \ijon to reach more state-dependent code. The seeds are valid but only execute few state transition sequences. \lf seems to be stuck in "shallow" states, and \ijon explores more state space with state tracing, while \csfuzz explores much more states that reach much "deeper" by both state tracing and heuristic algorithms. More discussions about our algorithm's contribution to the branch coverage can be found in \autoref{sec:sensitivity} and the growth trend of branch coverage can be found in \autoref{sec:growtrend}.

One exception is the \emph{Gstreamer}. We explain the reason with the huge state space in \emph{Gstreamer} which corresponds to a small code space. The enum-type variable \texttt{best\_probability}, which ranges from 1 to 100, represents the certainty about the media type of that stream. When \emph{Gstreamer} receives a stream, it gradually updates the value of \texttt{best\_probability} as more information becomes available. This state variable has 100 states and is updated frequently to seemingly random values, so \csfuzz generates much more inputs to explore the state space despite not achieving much more branch coverage. To confirm our hypothesis, we removed \texttt{best\_probability}, repeated this experiment, and saw a good improvement in the coverage results for \csfuzz. \lf would take 23 hours to achieve about the same coverage as \csfuzz achieves in 12 hours. 

\autoref{tab:coverage} (left) shows that, on average, \csfuzz achieves 2.20\% more branch coverage than \lf, 4.41\% more than \ijon, and 38.73\% more than \aflnet. \autoref{tab:coverage} (right) shows that \csfuzz achieves the same branch coverage 2.3x faster than \lf in the 23 hours, on average. For state-dependent code, this is either because \csfuzz takes much shorter time than \lf to reach the same coverage, (Time-to-coverage), or \csfuzz reaches parts of the code that \lf has been unable to reach (Branch coverage and RQ2 demonstrated in \autoref{fig:code_snippet}).
Looking at RQ1 and RQ2 together, there is a greater increase in transition coverage than in branch coverage. This further supports our hypothesis that {\em conventional branch coverage feedback is insufficient to measure the state space coverage of programs.}

We recall our example at \autoref{sec:motivating} to explain the branch coverage gap between \csfuzz and \lf. The function sequences \textcircled{1}\textcircled{2}\textcircled{3} and \textcircled{3}\textcircled{1}\textcircled{2} would result in the same branch coverage, but they are different regarding the value of the state variable (\texttt{stream->state}). This difference is captured by the STT in \csfuzz and hence, both two corresponding seeds are saved for further exploration, while \lf is unable to distinguish between these function call sequences and hence only the first input seed that result in one of these function calls will be saved.  Consequently, as the function \textcircled{4} is easier to be explored from mutating the seed resulting to \textcircled{1}\textcircled{2}\textcircled{3} than from the other seed, \lf has a high chance of missing the corresponding precious seed and becoming unable to cover the function \textcircled{4}. In our experiments, the function \textcircled{4} is always explored by \csfuzz faster than \lf, or is unexplored for \lf in 23 hours. There are more state-dependent code that produces the branch coverage gap.

\result{On average, \csfuzz achieves 2.20\% more branch coverage than \lf, 4.41\% more than \ijon, and 38.73\% more than \aflnet. \csfuzz achieves the same branch coverage about 2x faster than \lf.}

\subsection*{RQ.3 State Identification Effectiveness}\label{sec:stateideneff}
To evaluate how well our heuristic is able to identify true state variables, we manually examined the valid identified variables of enumerated type for all subjects as well as the corresponding Request for Comments (RFC) documents, which contain technical specifications and organizational notes for the Internet. 
To evaluate \emph{false positives}, we classified the identified variables into 1)~\texttt{configuration variables} if their values are got from static sources (configuration files, command-line parameters), 2)~\texttt{state variables} if their values are got from inputs and affect the program execution by the \texttt{if}- or \texttt{switch}-statements, 3)~\texttt{error or response codes} if their values are assigned as the return value of functions.
To evaluate the \emph{impact on state space exploration}, we measured the percentage of nodes in the STT that are actually state variables.
To evaluate \emph{false negatives}, we checked if the STT covers the protocol states as discussed in the corresponding RFCs.
Specifically, we checked a) if an RFC exists, b)~if the specified state machine is represented as enumeration type in the protocol implementation, and c) if STT covers these variables.

\begin{table}[]
\setlength{\tabcolsep}{4pt}
    \centering
    \footnotesize
    \begin{tabular}{@{}r|rrr|rrr|r@{}}
    \multirow{2}{*}{\textbf{Subject}} & \multicolumn{3}{c|}{\textbf{Enum}} & \multicolumn{3}{c|}{\textbf{STT}} &\\
    & All & State & Percentage & All & State & Percentage & RFC \\
        \hline
H2O	       & 43	    & 22	&   51.16\%	&   6418	&   6417	&   99.98\%  & \checkmark \\
MbedTLS    & 36	    & 33	&   91.67\%	&   167	    &   167	    &   100.00\% &  \checkmark	 \\
Curl       & 81	    & 44	&   54.32\%	&   35690	&   35629	&   99.83\%  & \checkmark	 \\
Gstreamer  & 218    & 151	&   69.27\%	&   11240	&   11224	&   99.86\%  &  NA	     \\
OpenSSL    & 64	    & 40	&   62.50\%	&   817	    &   789	    &   96.57\%  & \checkmark	\\
Live555    & 11	    & 10	&   90.91\%	&   17446	&   17446	&   100.00\% &  \ding{55}	 \\
OwnTone    & 51	    & 37	&   72.55\%	&   3671	&   3671	&   100.00\% &  NA	     \\
DCMTK      & 145    & 80	&   55.17\%	&   27178	&   27109	&   99.75\%  &  NA	    \\
        \hline
    \multicolumn{2}{c}{} & \textbf{Avg:} & 68.44\% & & \textbf{Avg:} & 99.50\% \\
    \end{tabular}
    \caption{[Left] The number of identified (All) and actual state variables (State) among enum-type variables. [Right] The number of nodes (All) and state-related nodes (State) in the STT constructed in 23 hours.}
    \label{tab:fp}
\end{table}

\emph{Presentation}. \autoref{tab:fp} shows the results of the state identification analysis. \emph{All} represents the number of enum-type variables or of nodes in STT. \emph{State} represents the number of state variables in enum-type variables or state nodes in STT. \emph{RFC} represents that if the protocol state machines stipulated in the Request for Comments(RFC) documents are covered by the STT. \emph{NA} represents \emph{Not Applicable}.

\emph{Results}. On average, 68.44\% of variables that were identified using our enum-based heuristic are actually state variables. However, the 32\% false positive rate has a low impact on the State Transition Tree (STT) construction. On average, 99.5\% of nodes in the STT are state variable values. The value of state variables change several times during the execution which is reflected in the growing STT.
The remaining 0.5\% of nodes are related to non-state variables. For instance, configuration variables are usually assigned once and often appear at the beginning of an execution (e.g., when the protocol exchange is configured). Error or response codes are usually written to log files or returned back to users \emph{once} per execution and without any change in value. This means non-state-variables rarely appear in the STT and thus do not affect our algorithm in \autoref{sec:fuzzing} or the achieved branch coverage.

\result{While an average of 32\% of identified variables are not actually state variables, an average 99.5\% of nodes in the STT constructed in 23 hours are referring to values of actual state variables. This is explained by the STT tracking \emph{changes} in variable values.}

We also investigated whether protocol states specified in the Request for Comments (RFC), if any, are reflected as enumeration-type state variables, and whether they appear in the STT. In other words, we investigated false negatives.
\emph{There is one false negative among our eight subjects.}
As we can see in the last column of \autoref{tab:fp}, three of the eight protocol implementations are proprietary protocols and no RFC exists, and the implementation (using state variables) already provides the ground-truth.\footnote{It is interesting to point out that these protocols cannot be tested by traditional specification-guided protocol testing tools.}  Four of the remaining five protocols implement the specified states as enumeration-type state variables.\footnote{\emph{H2O}(rfc7540), \emph{MbedTLS} (rfc8446), \emph{Curl} (rfc959, rfc3501, etc.), and \emph{OpenSSL} (rfc8446)}. For the Live555 implementation of the RTSP protocol, the corresponding RFC-2326 does specify the RTSP protocol state machine but it is implicitly represented by state variables related to data transport or parsing, which imply the protocol state machine. For instance, the enum-type variable \emph{fCurrentParseState} maintains the states of the media file parser, which must be the initial state under the \emph{READY} state of RTSP protocol and must not be the initial states under the \emph{PLAY} state of the RTSP protocol. However, as there is no explicit state variable, we count this as a false positive.

To evaluate state coverage regarding true state variables, we recorded the number of distinct values of true state variables in our experiments, as shown in \autoref{tab:valuecov}. On average, \csfuzz covers 88.98\% more states than \lf in 23 hours.
For \emph{Gstreamer}, which has the best improvement, there is a huge state space in which some states are easier to be explored, such as \texttt{best\_probability} variable (explained in RQ2).
For \emph{H2O}, which has least improvement, we found that the RFC-related variable \texttt{stream->state} only has eight possible values and all eight values are covered by all the fuzzers in 23 hours, and hence, there has been less room for state coverage improvement.  
In general, exercising new states is more challenging because specific inputs or external environment are usually needed, but \csfuzz still outperforms other fuzzers in state coverage.

We should also note that our automatic state identification is much more complete than a manual annotation, as in \ijon~\cite{ijonn}. For the \texttt{protocol-implementation-related example}\footnote{\url{https://github.com/RUB-SysSec/ijon-data/blob/master/tpm_fuzzing/src/CommandDispatcher.c\#L396}} in the paper, the authors annotated the message type in TPM to represent state (Listing.D\&Section.V.D). \ijon can generate more “distinct sequences” of message types. \csfuzz would identify the state variables \footnote{\url{https://github.com/RUB-SysSec/ijon-data/blob/master/tpm_fuzzing/src/CommandAudit.c\#L100}} of each TPM subsystem and track the state \footnote{\url{ https://github.com/RUB-SysSec/ijon-data/blob/master/tpm_fuzzing/src/Global.h\#L369-L374}} of these subsystems. We examined the experiment data of \ijon and found that the message sequences with the same message types may correspond to different subsystem states. For example, for the same message type sequence 32 and 34 (integers used in TPM to mark message type), the subsystem's state may be \texttt{SU\_RESTART} or \texttt{SU\_RESET}. It is because the subsystem's states are affected by previous execution results, not the message type. This shows how \csfuzz can automatically annotate state in a more fine-grained manner than human.

\begin{table}
    \centering
    \footnotesize
    \begin{tabular}{@{}r@{ \ }|@{ \ }r@{ \ }r@{ \ }r@{ \ }rr@{}}
        \textbf{Subject} & \scriptsize\textbf{\aflnet} &\scriptsize\textbf{\lf} & \scriptsize\textbf{\ijon} & \scriptsize\textbf{\csfuzz} & \multicolumn{1}{c@{}}{\textbf{Improvement}}\\\hline
H2O        & -	&   12	&   12	&   12	&   +0.00\% \\
MbedTLS    & -	&   14	&   17	&   21	&   +50.00\% \\
Curl       & -	&   55	&   67	&   75	&   +36.36\% \\
Gstreamer  & -	&   23	&   31	&   113	&   +391.30\% \\
OpenSSL    & 40	&   40	&   44	&   47	&   +17.50\% \\
Live555    & 15	&   15	&   17	&   19	&   +26.67\% \\
OwnTone    & 10	&   18	&   23	&   27	&   +50.00\% \\
DCMTK      & 6	&   5	&   7	&   12	&   +140.00\% \\
\hline
\multicolumn{3}{c}{ } &\multicolumn{2}{r}{\textbf{Avg: }} &\textbf{+88.98\%} \\
    \end{tabular}
    \caption{The average number of distinct state variable values observed in campaigns of length 23 hours (for enumeration variables that we know for sure represent states).} 
    \label{tab:valuecov}
\end{table}
 
\subsection*{RQ.4 Bug Finding Effectiveness}

Among known bugs from OSS-Fuzz and ProFuzzBench, we selected seven stateful bugs to evaluate if \csfuzz exposes stateful bugs faster, and three stateless bugs to evaluate if \csfuzz hinders the finding of stateless bugs. In ProFuzzBench, we selected the stateful bug CVE-2019-7314 in \emph{Live555} which was found by \aflnet. In OSS-Fuzz, we used the bug tracker that contains fuzzer-generated bug reports to chose the other seven bugs. Among \emph{stateful bugs}, we take the bugs that require at least two state transitions to be exercised before they can be exposed.
\emph{Stateless bugs} are the bugs that can be executed without any state transition.

\begin{table}[]
{\scriptsize
\setlength{\tabcolsep}{2.3pt}
\begin{tabular}{rr|cc|cc|cc|cc|r}
\multirow{2}{*}{\textbf{Subject}} & \multirow{2}{*}{\textbf{Bug}} &  \multicolumn{2}{c|}{\textbf{\aflnet}} & \multicolumn{2}{c|}{\textbf{\lf}} & \multicolumn{2}{c|}{\textbf{\ijon}} & \multicolumn{2}{c|}{\textbf{\csfuzz}} & \multirow{2}{*}{\textbf{Factor}} \\
   &   & Avg & Num & Avg & Num & Avg & Num & Avg & Num \\
\hline\hline
Stateful:\\
\hline
H2O         & 12096 & -     & -     & 4.17h & 12    & 4.89h  & 18   & 2.20h & 19    & 1.9 \\
OpenSSL     & 4528  & -     & -     & >23h  & 0     & 14.28h &  6   & 7.09h & 10    & $\infty$  \\
Curl        & 16907 & -     & -     & 6.29h & 19    & 4.92h &  11   & 1.98h & 19    & 3.2 \\
Gstreamer   & 20912 & -     & -     & 0.07h & 20    & 0.07h & 20    & 0.03h & 20    & 2.3  \\
Live555     & CVE   & 7.78h & 20    & 0.05h & 20    & 0.05h & 20    & 0.05h & 20    & 1.0 \\
H2O         & 3303  & -     & -     & >23h  & 0     & >23h  & 0     & >23h  & 0     & - \\
Curl        & 22874 & -     & -     & >23h  & 0     & >23h  & 0     & >23h  & 0     & - \\
\hline
\multicolumn{8}{c}{} & \multicolumn{3}{r}{\textbf{Avg: } 2.1x}   \\

Stateless:\\
\hline
H2O & 554   & - & - & 4.69h & 20 & 4.48h & 16 & 3.86h & 20 & 1.2 \\
Curl & 17954 & - & - & 0.18h & 20 & 0.21h & 20 & 0.15h & 20 & 1.2 \\
OpenSSL & 3122 & - & - & >23h & 0 & >23h & 0 & >23h & 0 & - \\
\hline
\multicolumn{8}{c}{} & \multicolumn{3}{r}{\textbf{Avg: } 1.2x}   \\

\end{tabular}}
\caption{Time to expose existing bugs: 20 campaigns, 23 hr.} 
    \label{tab:bug}
\end{table}

\emph{Presentation.} \autoref{tab:bug} shows the average time to expose all bugs in 20 campaigns of 23 hours. \emph{Num} is the number of campaigns in which the bugs were exposed. \emph{Factor} represents the factor improvement of \csfuzz to \lf based on the average time-to-bug. \emph{Bug} shows the bug identifier in the OSS-Fuzz issue tracker (for Live555, CVE-2019-7314).

\emph{Results.} For stateful bugs, \csfuzz is never slower than the competition in exposing the bugs. For OpenSSL, which has a more complicated state machine than others, \lf cannot trigger the bug in 23 hours, while \ijon and \csfuzz trigger it in 6 and 10 out of 20 fuzzing campaigns of 23 hours, respectively. For \emph{Live555}, \lf, \ijon, and \csfuzz expose the bug much faster than \aflnet because of the high throughput. In general, \ijon outperforms \lf, but it is  worse than \csfuzz. These known stateful bugs were originally found by \lf because few state transitions are needed (cf. \autoref{appendix:bugprev}).

For stateless bugs, \csfuzz shows a slight advantage over \lf and \ijon. We found that these stateless bugs are usually in the functions which are executed together with, but do not depend on state transitions, such as message parsing functions must be executed before state transitions (bug \#554). Our algorithms explore the state space more efficiently, facilitating the stateless bug finding as well.  

\result{On average, \csfuzz exposes stateful bugs about twice (2x) as fast as \lf and \ijon, and about 155x faster than \aflnet.}

\subsection*{RQ.5 New Bugs Found}\label{sec:newbugs}

\begin{table*}[htbp]
    \centering
    \footnotesize
    \begin{tabular}{r|cccccc}
        \textbf{Subject} & \textbf{version} & \textbf{Type} & \textbf{Stateful} & \textbf{State Variable} & \textbf{CVE} \\\hline
        Live555 & 1.08 & Stack-based overflow in liveMedia/MP3FileSource.cpp & True & Implicit & CVE-2021-38380 \\ 
        Live555 & 1.08 & Heap use after free in liveMedia/MatroskaFile.cpp & True & Explicit & CVE-2021-38381 \\ 
        Live555 & 1.08 & Heap use after free in liveMedia/MPEG1or2Demux.cpp  & True & Explicit & CVE-2021-38382 \\ 
        Live555 & 1.08 & Memory leak in liveMedia/AC3AudioStreamFramer.cpp & True & Explicit & CVE-2021-39282 \\ 
        Live555 & 1.08 & Assertion in UsageEnvironment/UsageEnvironment.cpp & True & Explicit & CVE-2021-39283 \\ 
        Live555 & 1.08 & Heap-based overflow in BasicUsageEnvironment/BasicTaskScheduler.cpp & True & Explicit & CVE-2021-41396 \\ 
        Live555 & 1.08 & Memory leak in liveMedia/MPEG1or2Demux.cpp & True & Implicit & CVE-2021-41397 \\ 
        OwnTone & 28.2 & Heap use after free in src/misc.c & False & - & CVE-2021-38383 \\ 
        DCMTK & 3.6.6 & Memory leak in dcmnet/libsrc/dulparse.cc & False & - & CVE Requested \\
        DCMTK & 3.6.6 & Memory leak in dcmnet/libsrc/dulparse.cc & True & Implicit & CVE Requested \\
        DCMTK & 3.6.6 & Heap use after free in dcmqrdb/libsrc/dcmqrsrv.cc & True & Explicit & CVE Requested  \\
        DCMTK & 3.6.6 & Heap-based overflow in dcmnet/libsrc/diutil.cc & True & Explicit & CVE Requested \\
        
    \end{tabular} 
    \caption{The 0--day security vulnerabilities found by \csfuzz. Eight CVEs were assigned. Four CVEs are under application.} 
    \label{tab:newbugs}
\end{table*}

We found 12 unique bugs in three of four subjects from ProFuzzBench. All new bugs are listed in \autoref{tab:newbugs}.  All of these bugs can be exploited remotely and have serious safety risks. So far, we have received 8 CVE IDs (we are in the process of applying for CVEs for the remaining 4 bugs). Out of the 12 new bugs we found, 10 are stateful. Of those, 7 exist in the states represented by explicit state variables of enumerated type, while the other 3 stateful bugs result from aggregated implicit states (cf \autoref{sec:implicit}). Two stateful bugs (the first and tenth line in \autoref{tab:bug}) can also be triggered by \lf because they depend on implicit states. One stateless bug (the eighth line in \autoref{tab:bug}) can also be triggered by \lf. With the same state variables identified by \csfuzz, \ijon additionally expose two bugs (the second and third line in \autoref{tab:bug}). All 12 new-found bugs cannot be triggered by \aflnet in 23 hours.

\emph{Case study: CVE-2021-39283\footnote{fifth line in \autoref{tab:newbugs}.} in Live555} (stateful bug in explicit state).
This bug is exposed with the following minimal prefix of RTSP states $\langle INIT \implies READY\implies PLAY\implies READY\implies PLAY\rangle$ and requires the following state transition in the MPEG parser (i.e., enum-type state variable \emph{MPEGParseState}) $\langle PARSING\_PACK\_HEADER \implies PARSING\_PACK\_HEADER\rangle$, all of which are tracked in the STT. After the first SETUP command, the first PLAY command starts the parsing process and sets the parser state for the media file header to PARSING\_PACK\_HEADER. The second set of SETUP and PLAY commands stops and restarts the parsing process, setting the parser state again to PARSING\_PACK\_HEADER. However, the number of bytes read in the first round is not reset properly, such that in the second round the media file header is parsed incorrectly which incurs an assertion error.

\result{\csfuzz found 12 new bugs in three widely-used and well-fuzzed stateful programs within 23 hours fuzzing. 10 of 12 bugs are stateful bugs.}
\section{Related Work}

\emph{State Space Inference}.
The most closely related works employ learning algorithms to infer state machines. There are two different ways to infer the state machine from a protocol implementation.  
{\em Passive learning} based approaches~\cite{prospex,pulsar,autofuzz,passive:learning:model,veritas, yu2020sgpfuzzer} infer a state machine by analyzing a corpus of protocol messages  observed from network traffic. A fundamental restriction to this kind of approach is the inability to learn protocols that communicate over an encrypted channel. 
{\em Active learning} based approaches~\cite{fuzztls,mace} actively query a protocol implementation with generated message sequences and inference a state machine using Angluin’s $L^*$ algorithm. HVlearn~\cite{hvlearn} infers DFA-models of SSL/TLS hostname verification implementations via learning algorithms.
In contrast, we do not infer the state machine from a protocol implementation but derive the STT structure from execution traces to identify the state space (Sec~\ref{sec:inference}). The construction of the STT requires no analysis of protocol messages or learning algorithms. We provide a new perspective of stateful system testing, in which protocol state transitions can be identified by changes on state variables.

\emph{Blackbox and Whitebox Protocol Fuzzing}.
{\em Blackbox fuzzers}~\cite{sulley, aspfuzz,tfuzz,protos,restler,peach, bestorm, codenomicon,boofuzz} take the protocol server under test as a black box and keep generating message sequences to the server for discovering bugs or security flaws.  Despite being simple and easy to be deployed, blackbox fuzzers are completely unaware of the internal structure of the program, perform random testing, and are thus ineffective.
{\em Whitebox fuzzers} use program analysis and constraint solving to synthesize inputs that exercise different program paths. MACE~\cite{mace} combines concolic execution and active state machine learning for protocol fuzzing. MACE uses concrete execution and symbolic execution to iteratively infer and refine an abstract model of a protocol so as to explore the program states more effectively.  However, these techniques suffer from scalability issues arising from the heavy machinery of symbolic execution. 

\emph{Greybox fuzzers for protocols}. These fuzzers observe program states in the execution and use coverage- and state-feedback to guide input generation for discovering program states. IJON~\cite{ijonn} uses human code annotations to annotate states. INVSCOV~\cite{balzarotti2021use} uses the likely invariants to partition program states. AFLNet~\cite{aflnet} uses response codes as states and evolves a corpus of protocol message sequences based on coverage- and state-feedback from executed sequences. 
For each run, AFLNet selects one of the states and fuzzes the entire sequence of messages which reach the state. To improve performance, \cite{chen2019exploring} uses fork mechanism to snapshot the states which are under fuzzing to avoid executing the whole protocol messages from the start. However, the fork points need to be specified manually in the code which demands the knowledge of the protocol state machine. 
Our approach comes under {\em greybox} fuzzing. \csfuzz does not require knowledge of the protocol state machine. Instead, we use state variables to identify the explored state space. In contrast to AFLNet, \csfuzz does not identify states by response codes. 
\section{Discussion}

The contribution of our work is in fuzzing reactive and stateful software systems. We propose to capture the exercised state space using a highly efficient dynamic construction of a state transition tree (STT). We developed a greybox fuzzer \csfuzz for stateful systems. Our experimental results show that \csfuzz discovered stateful bugs about twice as fast as \lf, the baseline greybox fuzzer that \csfuzz was implemented into. \csfuzz covers 33x more sequences of state transitions than \lf, 15x more than \ijon, and 260x more than \aflnet. At the same time, \csfuzz achieves the same branch coverage about 2x faster than \lf. Finally, \csfuzz found 12 new bugs.
Our tool \csfuzz  has been released open source.
\begin{center}
\textcolor{blue}{\url{https://github.com/bajinsheng/SGFuzz}}
\end{center}
%The experimental data is available from the anonymous link: \begin{center}\textcolor{blue}{\url{https://zenodo.org/record/5555955}} \end{center}
We followed a responsible disclosure policy for the bugs we found, and all reported bugs in this paper have now been fixed by developers.
%where we privately notified the project maintainers about the bug and how to reproduce it. Once the bug was fixed, we applied for a CVE identifier which comes with a security advisor to track the vulnerability. All reported bugs have been fixed.
We feel that tools like \csfuzz have significant practical usage since the security of the internet-facing protocol implementations is of paramount importance. 
%\csfuzz also opens up the possibility of dynamic specification inference \cite{Larus02} from fuzzing runs. 

%Dynamic specification mining \cite{Larus02}, where specifications are reverse-engineered from system execution traces, is a well-explored area. However, specification mining is primarily used for coarse-grained program comprehension, and is not directly used for finding bugs. Our work seeks to summarize stateful behavior during the fuzz campaign, while at the same time finding vulnerabilities in stateful software implementations. In future, one can extend fuzzers like \csfuzz to check more complex oracles or properties (say finding violations of temporal logic properties) against protocol implementations. Such extensions of \csfuzz would resemble software model checking, without incurring the large memory overheads of model checking that come due to state space caching. 

\bibliographystyle{plain}
\bibliography{references}

\appendix

\section{Appendix}
Due to the space limit, we present some of the evaluation results here.

\subsection{Sensitivity Analysis}\label{sec:sensitivity}

We conducted sensitivity analysis to evaluate the effectiveness of each heuristic in \autoref{sec:fuzzing}. To do this, we configured three new fuzzers at the \textsc{FuzzBench} framework: a)~\scx, a specific configuration of \csfuzz that does not have the algorithms of energy schedule in \autoref{sec:power} and prioritizing input bytes in \autoref{sec:mutation}. b)~\scy, another configuration of the \csfuzz that does not have prioritizing input bytes in \autoref{sec:mutation}. c)~\hf, a stateless fuzzers similar to \lf. The experiments were carried out with the identical configurations as our main experiments: 23 hours and 20 runs.

\autoref{tab:sensstatecov} shows the average state transition coverage. Similar to \lf, \hf is a stateless fuzzer, so its state transition coverage is little and close to \lf. This demonstrates that other stateless fuzzers are inefficient in state-space exploration as well. From \scx, to \scy, until \csfuzz, the state transition increase factor to \lf are 4.43x, 21.95x, and 33.92x on average which represent the contributions of each algorithm in \autoref{sec:feedback}, \autoref{sec:power}, and \autoref{sec:mutation}, respectively. It demonstrates that our heuristic algorithms (\autoref{sec:power} and \autoref{sec:mutation}) promote around 33x state transition coverage. Looking together with \autoref{tab:transitions}, \scx, which only has STT feedback, still slightly outperforms \ijon. We explain that \ijon only records the last four states in a state sequence for high throughput, while we capture all states of a sequence in the STT. 

\autoref{tab:sensbranchcov} shows the average branch coverage in 23 hours and the time it takes \hf, \scx, \scy, and \csfuzz, respectively to achieve the same coverage that \lf achieves in 23 hours. \hf excels at subjects \emph{H2O}, \emph{Curl}, and \emph{OpenSSL} and fails miserably at \emph{MbedTLS} and \emph{OwnTone}. The reason for this is that \scx, \scy, and \csfuzz are all based on \lf, while \hf is an entirely different fuzzer. In general, \scx slightly performs worse than \lf as it introduces performance cost for state tracing, and our heuristic algorithms overcome the performance cost as in \scy and \csfuzz. For \emph{Gstreamer} which has a big state space in small code space, \scy performs better than \csfuzz as the latter puts too much effort on the state space exploration. Time-to-coverage improvement is substantially greater than branch coverage improvement, which is consistent with our major results in \autoref{tab:coverage}. 

\begin{table}[]
    \setlength{\tabcolsep}{3.2pt}
    \centering
    \scriptsize
    \begin{tabular}{r|c|cccc}
        \textbf{Subject} & \textbf{\hf} & \textbf{\lf} & \textbf{\scx} & \textbf{\scy} & \textbf{\csfuzz} \\
        \hline
H2O		    & 82.65   & 70.80   & 114.50 & 1693.45 & 1849.30 \\
MbedTLS		& 18.95   & 22.80   & 32.55  & 49.40   & 50.80   \\
Curl		& 144.60  & 150.25  & 456.70 & 8899.00 & 14630.80\\
Gstreamer	& 48.70   & 49.40   & 186.50 & 2399.10 & 4067.30 \\
OpenSSL     & 25.34   & 23.95   & 30.00  & 31.03   & 33.10 \\
Live555     & 179.65  & 184.15  & 594.33 & 825.25  & 1162.30 \\
OwnTone     & 49.71   & 46.40   & 675.45 & 748.32  & 930.15 \\
DCMTK       & 183.22  & 189.25  & 1233.45& 3755.05 & 6737.05 \\
    \hline
    \end{tabular}
    \caption{Average state transition coverage for 20 campaigns of 23 hours. }
    \label{tab:sensstatecov}
\end{table}

\begin{table}[]
    \setlength{\tabcolsep}{3.2pt}
    \centering
    \scriptsize
    \begin{tabular}{r|c|cccc}
        \textbf{Subject} & \textbf{\hf} & \textbf{\lf} & \textbf{\scx} & \textbf{\scy} & \textbf{\csfuzz} \\
        \hline\hline
    \multicolumn{6}{l}{\textsc{Branch Coverage:}} \\
    \hline
H2O	        & 3207.90	& 2879.20	& 2885.50	& 3042.70	& 3050.68 \\
MbedTLS	    & 11735.15	& 11929.60	& 11890.35	& 11902.05	& 12057.68 \\
Curl	    & 19987.30	& 19262.16	& 16665.80	& 19619.77	& 19942.39 \\
Gstreamer	& 63235.45  & 63280.06	& 61807.94	& 62551.20	& 61698.56 \\
OpenSSL     & 13003.46  & 12463.14  & 12598.24  & 12596.10  & 12610.50  \\
Live555     & 2197.85   & 2244.63   & 2223.15   & 2270.00   & 2278.10   \\
OwnTone     & 2104.13   & 2184.64   & 2124.75   & 2292.00   & 2312.60   \\
DCMTK       & 6087.33   & 6042.15   & 5992.25 & 6033.70 & 6100.50   \\
    \hline
    \multicolumn{6}{l}{\textsc{Time-to-Coverage:}} \\
    \hline
H2O         & 3.60h     & 23.00h    & 12.41h    &  7.16h    & 6.69h     \\
MbedTLS     & >23.00h   & 23.00h    & >23.00h   &  >23.00h  & 17.86h    \\
Curl        & 14.25h    & 23.00h    & 22.58h    &  13.88h   & 12.23h    \\
Gstreamer   & >23.00h   & 23.00h    & >23.00h   &  >23.00h  & >23.00h   \\
OpenSSL     & 5.54h     & 23.00h    & 15.44h    & 16.32h    & 10.08h  \\
Live555     & >23.00h   & 23.00h    & >23,00h   & 19.33h    & 15.86h  \\
OwnTone     & >23.00h   & 23.00h    & >23.00h   & 8.18h     & 5.04h   \\
DCMTK       & 20.11h    & 23.00h    & >23,00h   & >23.00h   & 14.40h  \\
    \hline
    \end{tabular}
    \caption{Average branch coverage for 20 campaigns of 23 hours (top), and the time it takes each fuzzer to achieve the same coverage that \lf achieves in 23 hours (0.25h accuracy; bottom)}
    \label{tab:sensbranchcov}
\end{table}

\subsection{Growth Trend of Branch Coverage}\label{sec:growtrend}

\begin{figure}
    \centering
    \includegraphics[width=0.5\textwidth]{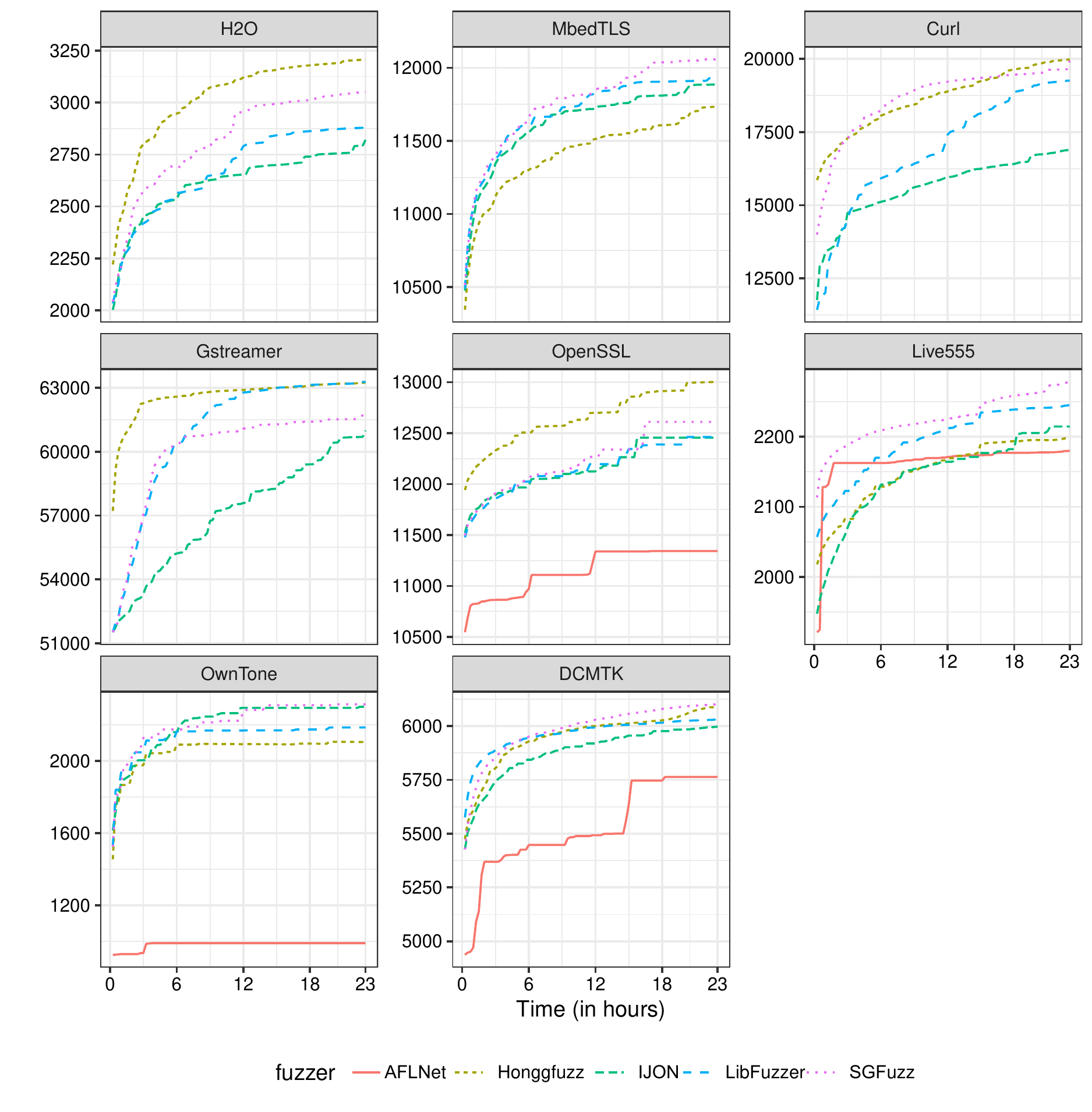}
    \caption{Average branch coverage over time for \csfuzz and \lf across all eight subjects in 23 hours.}
    \label{fig:coverage}
\end{figure}

\autoref{fig:coverage} shows the average branch coverage with time for all fuzzers in \autoref{tab:coverage} and \hf in \autoref{tab:sensbranchcov}. All fuzzers' branch coverage is nearly saturated after the 12 hours, although their state transition coverage is vastly varied. The slowing growth rate of branch coverage further explains why the time-to-coverage improvement is more significant than branch coverage improvement and branch coverage alone isn't enough to examine program state space. \aflnet is substantially slower than other fuzzers, especially on OwnTone. On OwnTone, \aflnet’s throughput is less than 10 inputs/second, while \csfuzz’s throughput is above 1000 inputs/second. Therefore \aflnet achieves much fewer branch coverage than others.

\subsection{Prevalence of Stateful Bugs}\label{appendix:bugprev}
To investigate the prevalence of stateful bugs, we studied the bugs that were reported in OSS-Fuzz against our subjects. We use the keyword \texttt{subject/fuzz\_driver status:Verified -Type=Build-Failure} to search available bugs on the OSS-Fuzz platform\footnote{\url{https://bugs.chromium.org/p/oss-fuzz/issues/list}}.
These bugs were reported against previous versions, and these reports are publicly available. A bug report links the bug fix and a crash-reproducing test case. We used this test case to debug these crashes and to determine whether these bugs are stateful. \autoref{tab:prevalence} illustrates the prevalence of stateful bugs. We classify a bug as \emph{stateful} if the crash location is reached after at least one state transition. We find that 57 (78.08\%) of all 73 bugs are stateful. 

\begin{table}
\centering\footnotesize
\begin{tabular}{@{ \ }r@{ \ }|@{ \ }r@{ \ }r@{ \ }l@{ \ }||@{ \ }r@{ \ }|@{ \ }rr@{ \ }l@{ \ }}
\textbf{Subject} & \textbf{Total} & \textbf{Stateful} & & \textbf{Subject} & \textbf{Total} & \textbf{Stateful}\\\hline
H2O & 9 & 6 & & Gstreamer & 34 & 28 & \\
OpenSSL & 4 & 3 & & Curl & 26 & 20 & \\
MBedTLS & 0 & 0 & &  &  &  & \\
\hline
\multicolumn{4}{c}{} & \textbf{Total:} & 73 & 57 & (78.08\%) \\
\end{tabular}
\caption{Prevalence of stateful bugs in OSS-Fuzz as of Sep'21.}
\label{tab:prevalence}
%\vspace*{-0.2in}
\end{table}

\result{Every four in five bugs that are reported in OSS-Fuzz for protocol implementations among our subjects are stateful.}

To understand the nature of bugs in protocol implementations, let us discuss all the bugs in H2O, our motivating example in \autoref{sec:motivating}. 

\emph{Stateful bugs in H2O}. All 6 stateful bugs in H2O happen after specific state transitions on the state variable \emph{stream->state} explained in \autoref{sec:motivating}. Each stream in HTTP2 protocol has a weight value to represent the priority of the stream to be handled and H2O defines a scheduling component to maintain it.
Bugs~\#12093, \#12096, \#12127, and \#12100 have the same root reason that they happen in the \emph{END\_STREAM} state of \emph{stream->state} within the scheduling component, so we only explain bug~\#12096 here. To trigger it, the H2O \emph{firstly} needs to set a higher priority for the current stream, (e.g., via the header frame or a dedicated priority frame), and \emph{then} close the stream  (e.g., via a "reset stream" or an "end stream" frame to trigger a scheduler relocation). At the same time, the value of \emph{stream->state} will be changed from \emph{IDLE} until \emph{END\_STREAM} representing a sequence of state transitions. The scheduling component only works for active streams, not closed streams. Lastly, when we assign a new priority to the closed stream, the bug~12096 is triggered because it does not check if the stream is closed.
Bug~\#2623 and \#3303 have the same root reason that they happen under the \emph{END\_STREAM} state of \emph{stream->state}, but not within the scheduling component. We explain the bug~\#2623 here. To trigger it, H2O \emph{firstly} receives a header frame with the end-of-stream attribute which indicates that no more frames for the stream. \emph{Then} the stream will be closed as the corresponding \emph{stream->state} is changed to \emph{END\_STREAM}. Lastly, when receiving another frame called trailing header for the closed stream, H2O may trigger the bug~\#2623 because of incomplete checking. 

\emph{Stateless bugs in H2O}. The 3 stateless bugs in H2O are observed before processing requests where no state transitions execute. Bug~\#2695 \#2923 happens in the response generation component. H2O defines an object called \emph{generator} to maintain related operations of response generation, but the \emph{generator} is used without checking if it is NULL incurring unexpected behaviors. Bug~\#37023 happens in the socket connection functions which is also executed before processing inputs without state transitions.

Totally, 6 out of 9 bugs in H2O are stateful. Although existing fuzzer can find stateful bugs, the stateful bugs they found only happen at the \emph{END\_STREAM} state of the \emph{stream->state} which represents the endpoint of the stream state. No bug is found during the state transitions except for the start and end states. It provides further evidence to support that the existing fuzzer cannot efficiently explore the state space of the protocol programs. On the other hand, even if the state space of the protocol program has not been explored enough, the number of stateful bugs is still greater than stateless bugs. Therefore, we claim that the stateful bugs are prevalent in protocol implementations. 

\subsection{Top-50 most widely used open-source protocol implementations.}\label{appendix:top50}

To investigate how states are tracked in the programs, We reviewed the code of top-50 widely used protocol implementations that are open-sourced. We used the same criterion as in \autoref{sec:stateideneff} to examine the state variables: the values of state variables are got from the input and affect the program execution by the switch or if statements. We only show the first state variables if several state variables exist in a subject.

\autoref{tab:top50} shows the results. It includes more than 16 protocols and 50 corresponding subjects. We marked the corresponding protocol of the state variable in the \emph{Protocols} column or noted \emph{Customized} if the state variables are used for customized states, such as connection and data processing states, not protocol states. Although some subjects (Sendmail, Postfix, e.g.,) are protocol implementations, the state variables we found in them are about customized states, not protocol states. This result explains that developers usually design many customized states in programs, and these states can not be found by protocol documents. The column of \emph{Variable Types/Values Examples} represent the state variable types: enumerated type or macro definition (\#define), and an example of state variable values. All 50 subjects implement their states by named constants, of which 44 subjects use enum-type variables.

\emph{Case study}. We used \emph{OpenSSL} to explain how the program states are manipulated by the corresponding state variables. \emph{Openssl} has a \textbf{handshake state machine} to maintain the state during the negotiation process between the two parties to establish a connection. OpenSSL defines fifty state values\footnote{\url{https://github.com/openssl/openssl/blob/5811387bac39cdb6d009dc0139b56e6896259cbd/include/openssl/ssl.h.in\#L1026}}, partially ordered, to implement it. We focus on the states on the server-side. At first, it is in the idle State~(TLS\_ST\_BEFORE). When receiving HELLO message from the client, it enters State~(TLS\_ST\_SR\_CLNT\_HELLO),  sends other optional messages one by one in States~(DTLS\_ST\_SW\_HELLO\_VERIFY\_REQUEST to TLS\_ST\_SW\_CERT\_REQ), and indicates it is done by State~(TLS\_ST\_SW\_SRVR\_DONE). Then when the client is ready and sends back optional messages, the server checks them in these States~(TLS\_ST\_SR\_CERT to TLS\_ST\_SR\_FINISHED, TLS\_ST\_SW\_CERT\_VRFY). At last, it sends final messages in States~(TLS\_ST\_SW\_SESSION\_TICKET to TLS\_ST\_SW\_ENCRYPTED\_EXTENSIONS) to confirm the server is ready too. States~(TLS\_ST\_SW\_KEY\_UPDATE, TLS\_ST\_SR\_KEY\_UPDATE) are used for updating keys after establishing the connection. States~(TLS\_ST\_SR\_KEY\_UPDATE, TLS\_ST\_SR\_END\_OF\_EARLY\_DATA) are about the early data feature introduced in TLS1.3 for the fast connection between client and server who share the same pre-shared key (PSK). 

\begin{table}[htpb]
    \centering\scriptsize
    \begin{tabular}{@{ \ }c@{ \ }|@{ \ }c@{ \ }|@{ \ }c@{ \ }}
        \hline
        \textbf{Protocols} & \textbf{Subjects} & \textbf{Variables Types/ Values Examples} \\
        \hline
        \multirow{3}{*}{\textbf{FTP}}   & Bftpd             & enum  / STATE\_CONNECTED      \\ 
                                        & LightFTP          & macro / FTP\_ACCESS\_NOT\_LOGGED\_IN      \\
                                        & vsftpd            & macro / PRIV\_SOCK\_LOGIN      \\\hline
        \textbf{SFTP}                   & ProFTPd           & macro / SFTP\_SESS\_STATE\_HAVE\_AUTH      \\\hline
        \multirow{6}{*}{\textbf{TLS/SSL}}&BoringSSL         & enum  / state13\_send\_hello\_retry\_request      \\
                                        & libssh2           & enum  / libssh2\_NB\_state\_idle       \\
                                        & MbedTLS           & enum  / MBEDTLS\_SSL\_HELLO\_REQUEST      \\
                                        & OpenSSL           & enum  / TLS\_ST\_BEFORE      \\
                                        & WolfSSL           & enum  / SERVER\_HELLOVERIFYREQUE.. \\ %ST\_COMPLETE      \\
                                        & Botan             & enum  / HELLO\_REQUEST              \\\hline
        \multirow{2}{*}{\textbf{SMTP}}  & Exim              & enum  / FF\_DELIVERED      \\
                                        & OpenSMTPD         & enum  / STATE\_HELO       \\\hline
        \multirow{10}{*}{\textbf{HTTP/2}}& aolserver        & enum  / SOCK\_ACCEPT      \\
                                        & H2O               & enum  / H2O\_HTTP2\_STREAM\_STATE\_IDLE      \\
                                        & httpd             & enum  / H2\_SS\_IDLE      \\
                                        & LiteSpeed httpd   & enum  / STREAM\_FIN\_RECVD      \\
                                        & NgHttp2           & enum  / NGHTTP2\_STREAM\_INITIAL      \\
                                        & Nginx             & enum  / NGX\_HTTP\_INITING\_REQUEST...       \\
                                        & Tengine           & enum  / NGX\_HTTP\_TFS\_STATE\_WRITE...       \\
                                        & thttpd            & macro / CHST\_FIRSTWORD       \\
                                        & WebSocket++       & enum  / READ\_HTTP\_REQUEST    \\
                                        & Lighttpd          & enum  / LI\_CON\_STATE\_DEAD      \\\hline
        \multirow{2}{*}{\textbf{RDP}}   & FreeRDP           & enum  / DRDYNVC\_STATE\_INITIAL      \\
                                        & xrdp              & enum  / WMLS\_RESET       \\\hline
        \textbf{NTP}                    & NTP               & macro / EVNT\_UNSPEC      \\\hline
        \textbf{IMAP}                   & Dovecot           & enum  / AUTH\_REQUEST\_STATE\_NEW      \\\hline
        \textbf{IRC}                    & UnreadIRCd        & enum  / CLIENT\_STATUS\_TLS\_STAR... \\\hline %TTLS\_HANDSHAKE      \\\hline
        \textbf{SMB}                    & Squid             & enum  / SMB\_State\_NoState      \\\hline
        \textbf{DAAP}                   & OwnTone           & enum  / PLAY\_STOPPED       \\\hline
        \textbf{SIP}                    & Kamailio          & enum  / H\_SKIP\_EMPTY      \\\hline
        \textbf{DICOM}                  & DCMTK             & enum  / DIMSE\_StoreBegin      \\\hline
        \multirow{2}{*}{\textbf{VNC}}   & libvncserver      & enum  / RFB\_INITIALISATION\_SHARED   \\
                                        & TigerVNC          & enum  / RFBSTATE\_UNINITIALISED      \\\hline
        \multirow{2}{*}{\textbf{RTSP}}  & ffmpeg            & enum  / RTSP\_STATE\_IDLE      \\
                                        & VideoLAN          & enum  / VLC\_PLAYER\_STATE\_STOPPED      \\\hline
        \multirow{1}{*}{\textbf{MQTT}}  & Mosquitto         & enum  / mosq\_ms\_invalid       \\\hline
        \multirow{14}{*}{\textbf{Customized}}& aria2        & enum  / STATUS\_ALL      \\
                                        & inetutils         & marco / TS\_DATA      \\
                                        & Gstreamer         & enum  / GST\_STATE\_READY      \\
                                        & OpenSSH           & enum  / MA\_START      \\
                                        & Pure-FTPd         & enum  / FTPWHO\_STATE\_IDLE      \\
                                        & libgcrypt         & enum  / STATE\_POWERON      \\
                                        & Postfix           & enum  / STRIP\_CR\_DUNNO      \\
                                        & Sendmail          & enum  / SM\_EH\_PUSHED      \\
                                        & Boa HTTPd         & enum  / READ\_HEADER      \\
                                        & WebServer         & enum  / STATE\_PARSE\_URI       \\
                                        & tevent            & enum  / TEVENT\_REQ\_INIT      \\
                                        & Live555           & enum  / PARSING\_VIDEO\_SEQUENCE...     \\
                                        & OpenMQTTGateway   & enum  / OFF \\
                                        & Curl              & enum  / MSTATE\_INIT      \\\hline
    \end{tabular}
    \caption{The top-50 most widely used open-source protocol implementations and their state variables.} 
    \label{tab:top50}
\end{table}

%%%%%%%%%%%%%%%%%%%%%%%%%%%%%%%%%%%%%%%%%%%%%%%%%%%%%%%%%%%%%%%%%%%%%%%%%%%%%%%%
\end{document}